\documentclass[prd,amsmath,amssymb, twocolumn]{revtex4-2}
\usepackage{graphicx}
\usepackage{dcolumn}
\usepackage{bm}
\usepackage{hyperref}
\usepackage{subfig}
\usepackage{caption}

\newcommand{\NMBH}{N_{\rm MBHB}}
\newcommand{\NUCB}{N_{\rm UCB}}
\newcommand{\order}[1]{\mathcal{O}(#1)}
\newcommand{\Sangria}{{\tt LDC2a-v2}}
\newcommand{\gbmcmc}{{\tt GBMCMC}}
\newcommand{\vbmcmc}{{\tt VBMCMC}}
\newcommand{\bayesline}{{\tt BayesLine}}
\newcommand{\mbhmcmc}{{\tt MBHBMCMC}}
\newcommand{\TheGlobalFit}{{\tt GLASS}}

\begin{document}

\title{Prototype Global Analysis of LISA Data with Multiple Source Types}
\author{Tyson B. Littenberg}
\affiliation{NASA Marshall Space Flight Center, Huntsville, Alabama 35811, USA}
\author{Neil J. Cornish}
\affiliation{eXtreme Gravity Institute, Department of Physics, Montana State University, Bozeman, Montana 59717, USA}
\date{\today}

\begin{abstract}
The novel data analysis challenges posed by the Laser Interferometer Space Antenna (LISA) arise from the overwhelmingly large number of astrophysical sources in the measurement band and the density with which they are found in the data. 
Robust detection and characterization of the numerous gravitational wave sources in LISA data can not be done sequentially, but rather through a simultaneous global fit of a data model containing the full suite of astrophysical and instrumental features present in the data.
While previous analyses have focused on individual source types in isolation, here we present the first demonstration of a LISA global fit analysis containing combined astrophysical populations.
The prototype pipeline uses a blocked Metropolis Hastings algorithm to alternatingly fit to a population of ultra compact galactic binaries, known ``verification binaries'' already identified by electromagnetic observations, a population of massive black hole mergers, and an instrument noise model.
The Global LISA Analysis Software Suite (\TheGlobalFit) is assembled from independently developed samplers for the different model components.
The modular design enables flexibility to future development by defining standard interfaces for adding new, or updating additional, components to the global fit without being overly prescriptive for how those modules must be internally designed.
The \TheGlobalFit\ pipeline is demonstrated on data simulated for the LISA Data Challenge 2b. 
Results of the analysis and a road-map for continued development are described in detail.
\end{abstract}

\maketitle

\section{Introduction}

The mHz band of the gravitational wave spectrum is expected to contain an unprecedented abundance of galactic, extra-galactic, and cosmological gravitational wave (GW) sources. 
The Laser Interferometer Space Antenna (LISA) will survey the mHz GW band and provide unique observational constraints on the formation and evolution of compact binaries in the Milky Way, the origin and growth of massive black holes throughout cosmic history, the dynamics of dense stellar environments in galactic nuclei, the fundamental nature of gravity and black holes, and more~\cite{LISA}.
The richness of the LISA source catalog comes at the price of a more complicated analysis framework than is required for currently operating GW observatories.
While aspects of the methodology developed for ground-based interferometers (many discrete sources)~\cite{GWTC-3} and pulsar timing (overlapping sources, sophisticated noise modeling)~\cite{Arzoumanian_2020} under-gird development of LISA analysis pipelines, new strategies are needed to account for the overwhelming number and density of GW signals in the LISA data.

The fundamental challenge of LISA analysis stems from the large number ($\mathcal{O}(10^4)$) and long duration (months to years) of detectable signals, resulting in non-negligible overlaps in time and frequency between discrete sources. 
As a result, analyses cannot treat sources independently and sequentially work through a list of candidate detections. Instead, the LISA analysis has to be approached globally, simultaneously fitting complete data models including all of the detectable GW sources and the detector noise. 
The need for a ``Global Fit'' was first described in 2005~\cite{Cornish:2005qw}, and has been identified as the primary challenge to the LISA analysis since early in the mission formulation~\cite{Vallisneri_2009}. This has lead to a coordinated effort to develop capable algorithms well in advance of mission operations~\cite{Babak_2010}.

Global fit analyses are not unique to LISA, as there are analogous methods used elsewhere in GW astronomy, and more broadly within astronomy and astrophysics (e.g. \emph{Gaia}~\cite{Gaia}). For LIGO-Virgo analysis, the {\tt BayesWave} pipeline simultaneously models Gaussian noise, non-Gaussian noise artifacts, and short-duration GW transients~\cite{PhysRevD.103.044006}. 
PTA analyses use a global fit to simultaneously model a correlated, stochastic gravitational wave background, a solar system ephemeris model, and multiple noise sources for each pulsar in the array~\cite{vanHaasteren:2008yh,Vallisneri_2020}. 
Some PTA analyses also perform a global fit for multiple source types, such as the signals from individual black hole binaries and a stochastic confusion noise from unresolved binaries~\cite{Becsy:2019dim}, or perform a {\tt BayesWave}-style analysis to reconstruct un-modeled burst signals~\cite{Becsy:2020utk}. 
Where the analogy ends is the scale of the LISA problem compared to elsewhere in GW astronomy, evident in the number of sources that are part of the global analysis, the diversity of source types (SMBHBs, EMRIs, UCBs, SGWBs, SOBHs, etc.), and data complications from multi-year integration times (glitches, nonstationary noise, gaps, etc.).

In this paper we present the first demonstration of a LISA global fit analysis contending with multiple source types.
We call the algorithm the Global LISA Analysis Software Suite, or \TheGlobalFit. 
As the proving ground for \TheGlobalFit\ we use the simulated data released in the second round of the LISA Data Challenges (Challenge \Sangria)~\cite{LDC2a}.  
The simulated data contain Gaussian detector noise; a simulated population of Milky Way ultra compact binaries (UCBs); 35 galactic UCBs already discovered by electromagnetic observations, the so-called ``verification binaries'' (VGBs); and a population of merging massive black hole binaries (MBHBs). 

The philosophy of \TheGlobalFit\ is to incorporate independently developed, stochastic sampling algorithms designed to address different facets of the full LISA analysis. 
\TheGlobalFit\ is then effectively an overarching umbrella that manages the interfaces, matches data formats, and orchestrates how the different samplers will work together to converge and adequately cover the target, high dimensional, joint posterior distribution function.   

\TheGlobalFit\ uses a four-component model to fit the simulated \Sangria\ data.  
The UCB, VGB, and MBHB models each employ analytic template waveforms to fit the detectable sources, while the noise level, including the unresolved astrophysical foreground from the galactic binaries, is fit with a phenomenological model.  
The parameters of the four model components are optimized using a blocked Metropolis Hastings (MH) algorithm. 
The test data set spans one year of simulated LISA observations however our analysis processes the data sequentially, analyzing increasingly large strides of the ``observed'' time series as we envision would be done during mission operations. 
Results from the analysis of the \Sangria\ data are compared to the input populations to assess the performance of the algorithm. 

\begin{figure}[htp]
    \centering
    \includegraphics[width=0.45\textwidth]{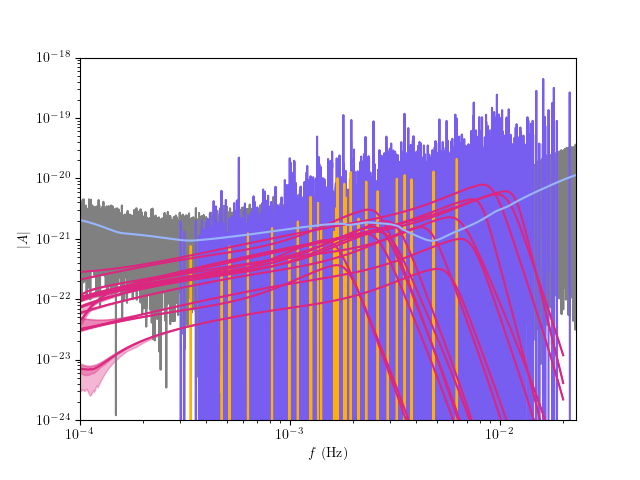}
    \caption{\small Median reconstructed global fit model components for the full 12 month \Sangria\ data, shown as the ASD in the TDI A channel. Gray is the residual, purple are the UCB detections, orange or the fits to the known binaries, magenta are the MBHB mergers, and light blue is the noise model.}
    \label{fig:reconstructions}
\end{figure}

Figure~\ref{fig:reconstructions} summarizes the \Sangria\ results by showing the combined reconstructed model components represented as the amplitude spectral density of the time delay interferometry (TDI) A channel~\cite{AET}.  The original data are shown in gray in the background of the figure. The purple lines are the posteriors of the reconstructed UCB waveforms while orange are the reconstructed VGBs. The magenta broadband curves are the reconstructed massive black hole signals, and the light blue curve is the noise model.  Note that at low frequency the credible 50 and 90\% credible intervals are visible in the black hole signals while otherwise the credible intervals on the reconstructions are too narrow to be visible in this plot. This figure represents the key result of this study: We are demonstrating a prototype pipeline able to simultaneously fit thousands of overlapping signals of different morphology and an \emph{a priori} unknown noise level. 

The remainder of this paper will provide a detailed description of the algorithm and a demonstration of the performance. 
Section \ref{sec:architecture} describes the analysis architecture, and how the different modules are integrated together into a global fit pipeline. Section \ref{sec:noise} describes the noise model and sampling algorithm, adapted from the \bayesline\ algorithm used for LIGO-Virgo noise modeling~\cite{PhysRevD.91.084034}. Section \ref{sec:gbmcmc} summarizes updates to the galactic binary sampler \gbmcmc\ first described in Ref~\cite{PhysRevD.101.123021}, while Section~\ref{sec:vbmcmc} specifies the configuration changes for \gbmcmc\ to perform targeted analysis of known binaries. Section~\ref{sec:mbh} describes how the \mbhmcmc\ massive black hole sampler from Ref~\cite{PhysRevD.105.044007} was adapted for this work. Section~\ref{sec:demo} presents the evolving results from 1.5, 3, 6, and 12 month analyses of the \Sangria\ data before we conclude in Section~\ref{sec:discussion} with a development road map to improve \TheGlobalFit's capabilities in order to analyze increasingly realistic LISA data.

\section{The \TheGlobalFit\ Architecture}\label{sec:architecture}
The central engine of \TheGlobalFit\ is a blocked Markov Chain Monte Carlo sampler~\cite{Andrieu:2003fk}. In the blocked sampling scheme, a subset of the model parameters (a block) is updated while holding all other parameters fixed.  Different blocks are updated independently in sequence, and the process cycles until the sampler has converged. For the LISA Global Fit problem, blocked MH samplers have two advantages. First, they work well for high dimension spaces when parameter correlations are confined to relatively small and \emph{a priori} identified sub spaces. Second, they are naturally modular, turning the daunting task of building an algorithm equal to the complexity of LISA data into a well defined set of components that are developed independently and then integrated.

The blocked MH scheme in \TheGlobalFit\ is hierarchical where the top level blocks, which we will refer to as ``modules,'' are the joint set of parameters for the different model components i.e.,  blocks for the noise, VGB, UCB, and MBHB parameter sets. The sampling within the VGB and MBHB modules is further grouped into blocks by individual sources, while the UCB module has one more layer of hierarchy--where model parameters are grouped by narrow-band frequency segments, and then by individual sources within each segment.

Each module uses a customized parallel tempered Markov Chain Monte Carlo sampler developed independently of one another.
The role of \TheGlobalFit\ is to coordinate which blocks are updating and to exchange state data between modules.
The modules work on their own subset of the data, use their own likelihood function, tempering scheme, proposals, etc., and in principle could even use a different representation of the data (e.g., time series, frequency series, or wavelets). 
In the current implementation of \TheGlobalFit\ each module is working in the frequency domain.

\begin{figure}[htp]
    \centering
    \includegraphics[width=0.45\textwidth]{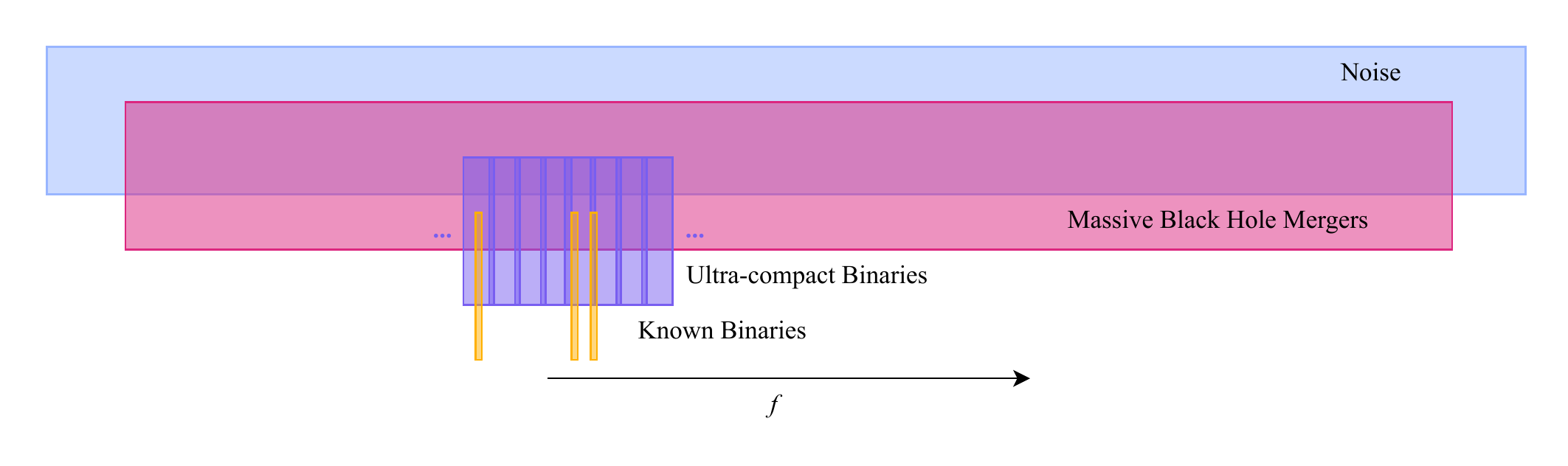}
    \caption{\small Schematic block diagram for how frequency domain data are segmented by the different model components.  The noise module must cover the full frequency range (light blue).  The MBHB module is broadband, covering almost the same width as the noise model (magenta). The UCB module divides the data into narrow-band, overlapping, segments (purple), while the VGB model targets only the frequency range spanned by each individual known binary (orange).}
    \label{fig:data_segments_schematic}
\end{figure}

Figure~\ref{fig:data_segments_schematic} is a schematic diagram for how modules operate on different bandwidths of the data. 
Note that the colors indicating each module will be consistent throughout this paper.
The noise model is fit over the full frequency measurement band of the data (light blue). 
For the \Sangria\ data we take that to range from ${\sim}10^{-5}$ to ${\sim}30$ mHz. 
Massive black hole mergers are broadband signals where the maximum frequency is determined by the total mass of the binary. 
The MBHB module bandwidth is dynamically based on the source parameters, but generally extends up to a few to $O(10)$ mHz (magenta).
The UCBs are narrow-band signals, generally spanning $\lesssim 10\ \mu{\rm Hz}$, but are by far the most numerous of the LISA sources, and will be found throughout the measurement band of our analysis, though sparsely above 10 mHz. 
The UCB module consists of several instances of the same sampler, each focusing on a band-limited segment of data (purple).  
The bandwidth of each segment depends on the frequency, using smaller segments where sources are most densely spaced.
Finally the VGB module is conducting a narrow-band targeted analysis for individually known sources (orange).

To understand the interface between modules consult the joint likelihood function for the global fit:
\begin{widetext}
\begin{equation}\label{eq:likelihood}
p(d|\theta) = (2\pi)^{-\frac{N}{2}} \det \left(C(\theta_{\rm noise})\right)^{-\frac{1}{2}} e^{-\frac{1}{2} \left( d - h(\theta_{\rm MBHB}) - h(\theta_{\rm UCB}) - h(\theta_{\rm VGB}) \right )^T C(\theta_{\rm noise})^{-1} \left( d - h(\theta_{\rm MBHB}) - h(\theta_{\rm UCB}) - h(\theta_{\rm VGB}) \right ) }
\end{equation}
\end{widetext}
where $d$ is the data, $N$ is the number of data points, $C$ is the noise covariance matrix, $\theta$ represents the full parameter set, $\theta_i$ are the model parameters in the block for module $i$, and $h$ are the co-added detector responses to the modeled sources in each module. 
For example, $h(\theta_{\rm MBHB})$ is really shorthand for $\sum_n h(\theta_{\rm MBHB}^n)$ where $n$ is indexing all of the sources in the MBHB model. 

Sticking with the MBHB example, the sampler adopted for that model was developed assuming no other sources in the data, and a known noise covariance matrix. 
The internal likelihood for the $k^{\rm th}$ component of the MBHB module is then just 
\begin{equation}
    p(d|\theta_{\rm MBH}^k) = e^{-\frac{1}{2} \left( \bar{d} - h(\theta_{\rm MBHB}^k) \right )^T C^{-1} \left( \bar{d}- h(\theta_{\rm MBHB}^k)\right ) }
\end{equation}
where the normalization term is absent because the sampler only considers the likelihood ratio between two points in parameter space and the covariance matrix is independent of the (MBHB) parameters.
To interface this sampler with the rest of \TheGlobalFit\, at the beginning of each one of the MBHB blocks' updates the covariance matrix is replaced based on the current state of the noise model $\theta_{\rm noise}$ and the ``data'' as seen by the MBHB sampler is the \emph{residual} after subtracting all other model components--as far as the MBHB sampler for the $k^{\rm th}$ MBHB is concerned $\bar{d} = d - h(\theta_{\rm UCB}) - h(\theta_{\rm VGB})-\sum_{n\neq k} h(\theta_{\rm MBHB}^n).$ 
The MBHB sampler is otherwise blissfully unaware of what is happening in the rest of the global fit.  
It is the job of \TheGlobalFit\ to keep track of the current state of each module, prepare the effective data and noise covariance matrix, and refresh the likelihood (using the new effective data and noise model) of the current state before a sampler updates it's block of parameters.
An identical argument applies to the other modules.

\begin{figure*}[htp]
    \centering
    \includegraphics[width=0.8\textwidth]{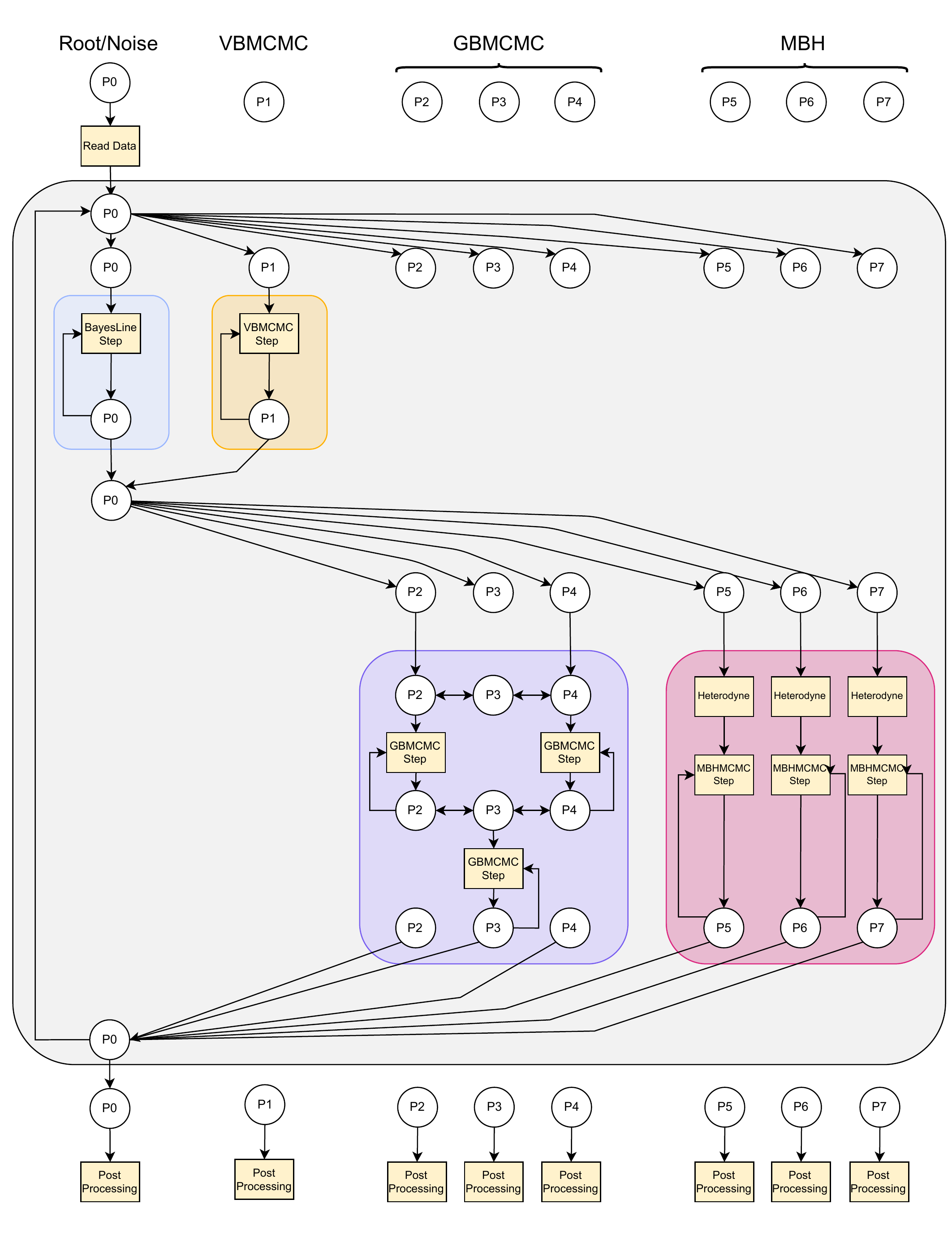}
    \caption{\small Block diagram of Global Fit architecture. Process P0 is the root process and runs the noise model sampler. Processes P1 to P3 are for the UCB model.  Processes P4 to P6 run the MBHB model. Gray is the blocked MH sampler.  Purple are the $\NMBH$ independent MBHB sampling steps.  Green are the $\NUCB$ coupled UCB processes, which exchange only between adjacent segments.  Orange is the Noise model which is run on the root processes. Data from all processes are shared with, and broadcast from, root. In practice $\order{10^3}$ UCB processes are needed, and $\order{10}$ MBHB processes.}
    \label{fig:block_diagram}
\end{figure*}

The individual samplers for the modules have been developed and published elsewhere and will be briefly summarized in later sections before focusing on updates made to each of them for the LISA global analysis.
To understand how the data is shared between modules a block diagram of the workflow is shown in Fig.~\ref{fig:block_diagram}. 
The diagram is a simplified version of the true workflow, depicting only three UCB and MBHB nodes each.  In practice \TheGlobalFit\ uses several hundred UCB nodes and one MBHB node per source in the model.
The noise, VGB, UCB, and MBHB model updates are executed by the \bayesline, \vbmcmc, \gbmcmc, and \mbhmcmc\ blocks, respectively.
\TheGlobalFit\ uses the Message Passing Interface (MPI) standard to exchange information between the different modules. 
For shorthand we will refer each MPI process as a ``node'' of the analysis, though in practice we use multiple MPI processes per node.

The \bayesline\ module uses a single node which also serves as the root process responsible for the work shared by all nodes and the overall orchestration of the analysis (P0 in the flow chart). 
At start up the root node handles data parsing, selection, and conditioning before broadcasting the data and the initial state of each sampler to all other worker nodes.
During each iteration of the sampling nodes P0 and P1 first update their parameter blocks for the noise model and VGB model, respectively.  
At the end of the noise and VGB module updates (each involving several internal MCMC steps for each model), the VGB process sends the current state of the VGB model in the frequency domain to the root process.  
The root process then broadcasts the current state of the noise model and VGB model to the UCB (P2-P4) and MBHB (P5-P7) nodes.
The UCB and MBHB processes then create their respective residuals and update their block of parameters.

Each UCB process is responsible for a narrow-band segment of the data but care must be taken at the segment boundaries where individual sources can span the interface.  
Each node shares its current state with the neighboring nodes (i.e. the adjacent frequency segments) so that the receiving node can remove the state of the neighboring model when forming the residual that will effectively serve as the data for the current update. 
The segments overlap in frequency and each node is responsible for fitting to the sources in its half of the overlapping region.  
This overlap, which is set to be a factor of two larger than the typical bandwidth of a source at that frequency, ensures that the templates for sources located near the boundaries are not artificially truncated. 
To preserve the correlations between sources that are close to one another on either side of a boundary, the UCB nodes alternate which block is updating and which is waiting. 
For example, all of the odd-numbered processes will update their parameters, exchange with their neighbors, and then the even processes will update.  

For the MBHB modules, additional pre-processing is needed once their residuals are formed.  
The MBHB module relies on the heterodyned likelihood described in~\cite{Cornish:2010kf,Cornish:2021lje,Cornish:2021smq} and recomputes the coefficients based on the current state of the sampler, as well as updating proposal distributions used within the sampler that use the information matrix as an approximation to the covariance matrix of the posterior. 
After that pre-processing each MBHB module updates its parameter block in parallel with the other MBHBs in the fit. 
At the end of the UCB and MBHB updates each module sends the current state of its model in the frequency domain to the root process to broadcast to all of the other nodes, and the entire cycle repeats.  
After many iterations when the sampling has finished each process performs a minimal level of post-processing to prepare for the next stage of the pipeline when the posterior samples are consolidated into a source catalog.

Note that in the traditional blocked MH sampling scheme only one block of parameters are updated at a time while the others are held fixed whereas \TheGlobalFit\ is updating blocks of parameters in parallel.  
This was a choice made to improve the computational efficiency of the algorithm.  
The current bottlenecks for the analysis are the heterodyne step for the MBHB model and the convergence time for the highest frequency UCBs.  
Those two aspects of the problem set the scale for the cost of each iteration and the number of iterations needed, respectively. 
To maximize efficiency, we do enough MBHB parameter updates to match the cost of updating the heterodyne. 
The number of internal \gbmcmc\ updates per cycle of the full sampler are then dynamically adjusted to take approximately the same amount of computational time as the MBHB models. 
The comparative costs of the noise model and \vbmcmc\ updates are significantly lower, similar to the costs of the data sharing and common processing that must get done before each iteration (e.g. writing results to file, etc.). 
This scheme was thus created to maximize the duty cycle of individual nodes by minimizing the amount of time nodes are blocked waiting for other processes to finish their work.  
While technically violating the conditions needed to have the resulting samples be representative of the target posterior distribution function, the effects are only noticeable for blocks that are correlated. 
Updating alternating UCB blocks ensures that no correlated UCB parameters are being altered in parallel as the data segments for each UCB node are larger than the bandwidth of a single UCB signal. 
Similar arguments can be made between other blocks being updated in parallel although a production analysis on observational data requires more through testing for confirmation, and/or a more conservative approach and a higher computational cost.
It is a trivial rearrangement of where the MPI exchanges take place to revert to the traditional serial update of all parameter blocks.

Each of the samplers use parallel tempering~\cite{Swendsen:1986} to improve the mixing of the chains. 
Parallel tempering is especially critical for promoting transitions between models in the trans-dimensional samplers.
The tempering scheme is independently developed and tuned for the different modules, and only the zero temperature chain parameters or state are shared between different processes. 
Each of the parallel tempering samplers is multi-threaded ideally using a single CPU per chain.
In practice there is a trade space between the number of resources needed for the analysis and the amount of time those resources are needed.
Processing of the \Sangria\ data was done on Amazon World Service (AWS) cloud computing infrastructure which favors smaller-scale jobs running for longer. 
As a result the \Sangria\ analysis used multiple threads per CPU. 
The final configuration for the full \Sangria\ analysis (12 months of simulated data) used 624 MPI tasks and 12 CPUs per task for a total of 7488 CPUs.  
The run was deployed on $78$ $\times\ 96$ CPU nodes.  
The noise model and verification binary modules used one MPI task each.  There were 15 MBHB mergers in the data each run with a dedicated MPI task. 
The remaining 607 MPI tasks were dedicated to the UCB model covering $.03$ to $23$ mHz.

\section{Description of the individual samplers}
The individual model components that are integrated into the \TheGlobalFit\ architecture are independently developed, described, and published. 
Each is still under active development so it is useful to overview each sampler with emphasis on updates that have been made since the most recent publications.
\subsection{Global Noise Model}\label{sec:noise}
For the noise model we use an adaptation of the \bayesline\ algorithm originally developed for LIGO-Virgo noise modeling~\cite{PhysRevD.91.084034}.
The original \bayesline\ algorithm fits the power spectral density (PSD) of the noise $S_n(f)$ independently in each detector. 
The LIGO-Virgo version of the pipeline uses a two-component fit to phenomenologically model the noise spectrum.
The main component is a broadband noise spectrum that looks similar to a sum of power laws, with steeply rising noise at low frequency and a more gradual increase at high frequency. 
Modeling the actual LIGO-Virgo data with a broken power law is not sufficiently flexible so \bayesline\ uses a cubic spline interpolation where each spline control point $i$ is parameterized by its frequency and PSD level $[f^i,S^i_n]$.
The location of the spline points, as well as the total number, are free parameters sampled over with a trans-dimensional MCMC.
For the LIGO-Virgo application \bayesline\ also includes a linear combination of Lorentzians to fit the narrowband features in the spectrum due to calibration lines, the power supply, resonances of the mirror suspension system, etc.
The simulated LISA data do not contain narrowband noise features and so \TheGlobalFit's implementation of \bayesline\ does not use the line model.  
Note, however, that there were spectral lines in the LISA \emph{Pathfinder} data and so future version of the model will need such a feature~\cite{LISAPathfinder:2019eny}.
There are two important differences between the \bayesline\ implementation integrated into \TheGlobalFit\ and that which has been used for LIGO-Virgo data.

First, the LISA noise spectrum spans the frequency regime where finite arm length effects of the detector response are in the measurement band, unlike ground-based interferometers which operate entirely in the long wavelength limit.
The arm length manifests in the noise spectrum as sharp features where the PSD, and instrument response, formally go to zero for signals with wavelength that fit an integer number of cycles within the detector arm.
Mathematically this is a result of terms proportional to $\sin^2(f/f^*)$ appearing in the detector response functions with the ``transfer frequency'' $f^*\equiv c/2\pi L$ where $c$ is the speed of light and $L\approx2.5$ Gm is the arm length of the LISA detector. 
The resulting spectrum is not well modeled by a spline interpolation at high frequencies.
However, the difficult features for a spline interpolation to track are a purely geometric effect set by the size of the detector. 
We therefore model the \emph{difference} between a reference noise spectrum, including the geometric effects, and the observed data $S_n^{\rm modeled} = S_n^{\rm observed} - S_n^{\rm reference}$, i.e. we are fitting for broadband differences between the reference noise level, derived from the current best estimate of the LISA performance, and the observed data.  
Where the reference model is accurate the modeled PSD will be consistent with zero.  

Second, the interpolation between control points for the \TheGlobalFit\ application of the spline model employs Akima splines~\cite{10.1145/321607.321609} rather than the cubic splines used in the LIGO-Virgo applications. 
The Akima splines are less prone to oscillations between control points by relaxing the requirement of a continuity in the second derivative of the interpolated curve.
The tendency for cubic splines to oscillate is exacerbated by the large dynamic range and steeply changing spectrum at low frequency. 
Akima splines perform better on the LISA spectrum and are worth considering for LIGO-Virgo noise modeling as well.

\subsection{GBMCMC Updates}\label{sec:gbmcmc}
The UCB sampler is the \gbmcmc\ pipeline described in Ref.~\cite{PhysRevD.101.123021}.
The \gbmcmc\ application is the latest in a long line of algorithms designed for the LISA galactic binaries which partition the frequency domain data into many narrow-band segments and uses model selection to determine the number of detectable binaries in each segment~\cite{Crowder:2006eu,PhysRevD.84.063009}.
The model selection method of choice used by \gbmcmc\ is a transdimensional, or reversible jump Markov Chain Monte Carlo (RJMCMC).

Of the different samplers integrated by \TheGlobalFit\, \gbmcmc\ has seen the most additional development.
The sampler was updated to use multi-threading for the parallel tempered chains, making a significant improvement in the run time especially when leveraging the increasingly large number of CPUs available per node.

The pipeline has also updated the way results from previous runs are incorporated as proposal distributions for subsequent analyses.  
As described in Ref~\cite{PhysRevD.101.123021}, the LISA data are processed in increasingly-long time epochs, starting with the first 1.5 month segment of data and re-processing each time the available data has doubled (i.e., after 3, 6, and 12 months).
In the first version of the \gbmcmc\ pipeline multivariate Gaussian proposals were built using the covariance matrix of the posterior samples for each UCB in the catalog.
In the latest version the single Gaussian proposal was replaced by a Gaussian Mixture Model (GMM) which is fit using the Expectation Maximization (EM) algorithm run on the posterior samples for sources in the previous epoch's catalog.
Evaluating the GMM proposal is more computationally costly than the single multivariate proposal, but we have found it to be offset by the improvement in convergence time. 

The \gbmcmc\ sampler now also includes a basic ``split-merge'' proposal for trans-dimensional steps, whereas the original algorithm only used ``birth-death'' moves.  A birth-death move chooses to either remove or add a feature to the model (in \gbmcmc's case, a source from or to the fit). 
The split-merge proposal attempts to divide a single feature into two, or combine a pair of features into one. 
The current implementation of the the split-merge proposal is naive, choosing to remove one source and replace it by two drawn from the same distribution as is used by the birth-death moves, or to remove two of the current sources and replace them by a single draw.  
In other words, the current split-merge proposal is really two birth moves and one death move, or two death moves and one birth move, respectively. 
Further development of more efficient split-merge proposals will be a critical area to improve the sampling.

Finally, in the previous applications of the \gbmcmc\ sampler the frequency segments were of equal bandwidth over the entire observing band.  
Because each segment was analyzed independently the overall number of segments (and therefore nodes) needed for the analysis was not a limiting factor in its deployment.
Within the \TheGlobalFit\ architecture when all of the processes are communicating via MPI we need to be more parsimonious about the number of segments being analyzed. 
To that end we adopted an adaptive segment size depending on the source density.  
At low frequency where the source density is the highest the segments are more narrow, and at high frequency where the source density is low (and the signals have larger bandwidth) the segments are wider. 
The exact segmenting was fixed before the \Sangria\ analysis was started and kept the same for each epoch's analysis. 
A more efficient approach would be to use the previous epoch's catalog to dynamically determine where to best place the segment boundaries to both keep the source density per segment near constant, and to avoid having loud signals near segment boundaries as an insurance policy, even though the ``edge effects'' of the segmenting are already ameliorated by \TheGlobalFit's data sharing scheme between UCB segments.

\subsection{VBMCMC Updates}\label{sec:vbmcmc}
The verification binary sampler (\vbmcmc) in \TheGlobalFit\ is identical to \gbmcmc\ but is run in a different configuration.
Whereas \gbmcmc\ is performing a blind search for UCBs in the LISA data, part of which includes a model selection step to determine if a candidate source in the data is detectable, \vbmcmc\ is executing a targeted analysis of binaries which have already been identified as LISA sources by EM surveys~\cite{Kupfer_2018}.

The \vbmcmc\ sampler therefore uses a fixed-dimension analysis with priors on the orbital period and sky location of the binaries derived from the EM observations.  
Because the sky localization from the EM observations is orders of magnitude more precise than LISA will ever achieve, the sky location parameters in \vbmcmc\ are fixed to the EM-observed values, effectively using delta functions for the priors.
The same is true for the orbital period of the binaries (converted to GW frequency for the \vbmcmc\ model). 
While it is possible that some of the currently-known binaries' orbital period measurements are not as precise as what LISA will infer, the long temporal baseline of EM observations should effectively pin the orbital period for LISA observations assuming continued effort to periodically monitor the known binaries prior to LISA's launch.
For the known binaries where this is not the case, replacing the delta function prior on GW frequency with a Gaussian distribution with width from the EM uncertainties is a trivial change.

Using a fixed-dimension targeted search for the known binaries instead of retroactively extracting the known binaries from the full UCB source catalog allows for upper limits to be set on binary parameters (most notably the GW amplitude) in the event that some of the known binaries are below the detection threshold at the time of the global fit analysis, perhaps due to elevated levels of the astrophysical foreground (confusion) noise, or because they will require longer integration times with LISA before becoming detectable. 

The targeted \vbmcmc\ analysis will also reduce contamination of known binaries from other loud sources at similar orbital periods, as many of the currently know binaries are at frequencies where the galactic source density is expected to be highest.
Analysis of known binaries will be particularly vulnerable to source contamination early in the LISA mission when the frequency resolution and integrated signal to noise levels are still improving at the same time as when UCB observations may play an important role in the early phases of instrument characterization.

\subsection{MBHB Updates}\label{sec:mbh}
The MBHB module uses elements of the {\tt LISA-Massive-Black-Hole-Binary} pipeline originally developed for low-latency detection and parameter estimation of massive black hole mergers with LISA~\cite{Cornish:2021smq}. 

The pipeline starts with a pre-processing search phase that uses short stretches of data (typically a few weeks), treating the galactic foreground as a noise source, and using a maximized likelihood function to rapidly lock on to any massive black hole binaries. 
The search is repeated on each segment of data after subtracting the previously found signals until no additional sources above a S/N threshold are found.
The rapid search is then followed with a full MCMC exploration of each source, taken one at a time, that refines the parameter estimates. These estimates are then used as the starting point for the MBHB analysis in \TheGlobalFit.

The same PTMCMC sampling routine is used in the global fit, but now with the noise model replaced by the spline model, and with the resolved UCBs  subtracted. 
Another key difference is that the model components are updated in an alternating fashion, in contrast with the low latency analysis where the noise model is fixed and the MBHB signals are analyzed sequentially. 

The MBHB block of parameters is updated as follows: When it comes time to update a particular MBHB, the sampler receives the current state of the residuals  residual constructed from the other model components. That is, the original data with the current state of the combined UCB and VGB models, as well as the other MBHB waveform models, subtracted. 
A key element of the MBHB sampler is the use of heterodyning to accelerate the likelihood calculations~\cite{Cornish:2010kf,Cornish:2021lje,Cornish:2021smq}. 
The heterodyne is computed using a reference waveform--in this case one based on the parameters of the MBHB model at the end of the last update, and the current state of the residual. 
The computational cost of setting up the heterodyne is equal to a few times the cost of a standard likelihood evaluation, so to be cost effective it makes sense to perform hundreds of iterations with the fast heterodyned likelihood before moving on to the next block in the sequence of updates.

The current version of the MBHB sampler uses customized implementation of the {\tt IMRPhenomD} waveform model which describes the dominant $(2,2)$ harmonic for spin-aligned, quasi-circular binaries. 
In future versions of the sampler the waveform model will be generalized to include spin precession effects, sub-dominant waveform harmonics, and eventually, orbital eccentricity.

Another limitation of the current implementation is that the dimension of the MBHB model is fixed to whatever was found by the low latency search. 
The MBHB model also needs to be trans-dimensional with the results from the search phase being used to propose adding or removing sources.
 
\section{Demonstration}\label{sec:demo}
Having described the overall \TheGlobalFit\ architecture we now turn to a demonstration of the pipeline's performance on simulated LISA data.
The test data set was produced by the LISA Data Challenge (LDC) team and is the first of the LDC data sets to contain a combination of different source types~\footnote{During a previous iteration of the LISA mission the Mock LISA Data Challenges also included a multi-source-type data sets which were left un-analyzed}.
The following demonstration of \TheGlobalFit's current capabilities uses the \Sangria\ data which contain ${\sim30}$ million galactic UCBs, 37 VGBs, 15 MBHB mergers, and an unspecified instrument noise level. 
The \Sangria\ data span one year of LISA observation time assuming a 100\% duty cycle.
In reality there will be periodic and sporadic interruptions to the data taking which will require further development of \TheGlobalFit's noise model and likelihood functions.

For each LDC simulation there are ``blind'' and ``training'' data sets, where the training data contain the list of signals (injections) used for the simulation.
The blind data are simulated using a different realization of the same population that is found in the training data.
For this demonstration the training data are used, enabling assessment of the pipeline performance through comparisons of the resulting source catalog to the injected signals.
The \Sangria\ data TDI channels are generated assuming an equal arm interferometer with stationary instrument noise such that the TDI ``A'' and ``E'' channels are noise-orthogonal~\cite{AET}.  
The \TheGlobalFit\ noise model correspondingly uses an independent fit to the A and E instrument noise levels. 
This simplifying assumption will need to be relaxed for analysis of the observational data but will only effect the computational cost of the analysis by introducing correlations between TDI data streams which result in non-zero off diagonal terms in the noise covariance matrix at each frequency. 
The resulting likelihood evaluations require more operations to compute (see Eq.~\ref{eq:likelihood}) but will not effect the overall complexity of the global fit. 

\begin{figure}[htp]
    \centering
    \includegraphics[width=0.48\textwidth]{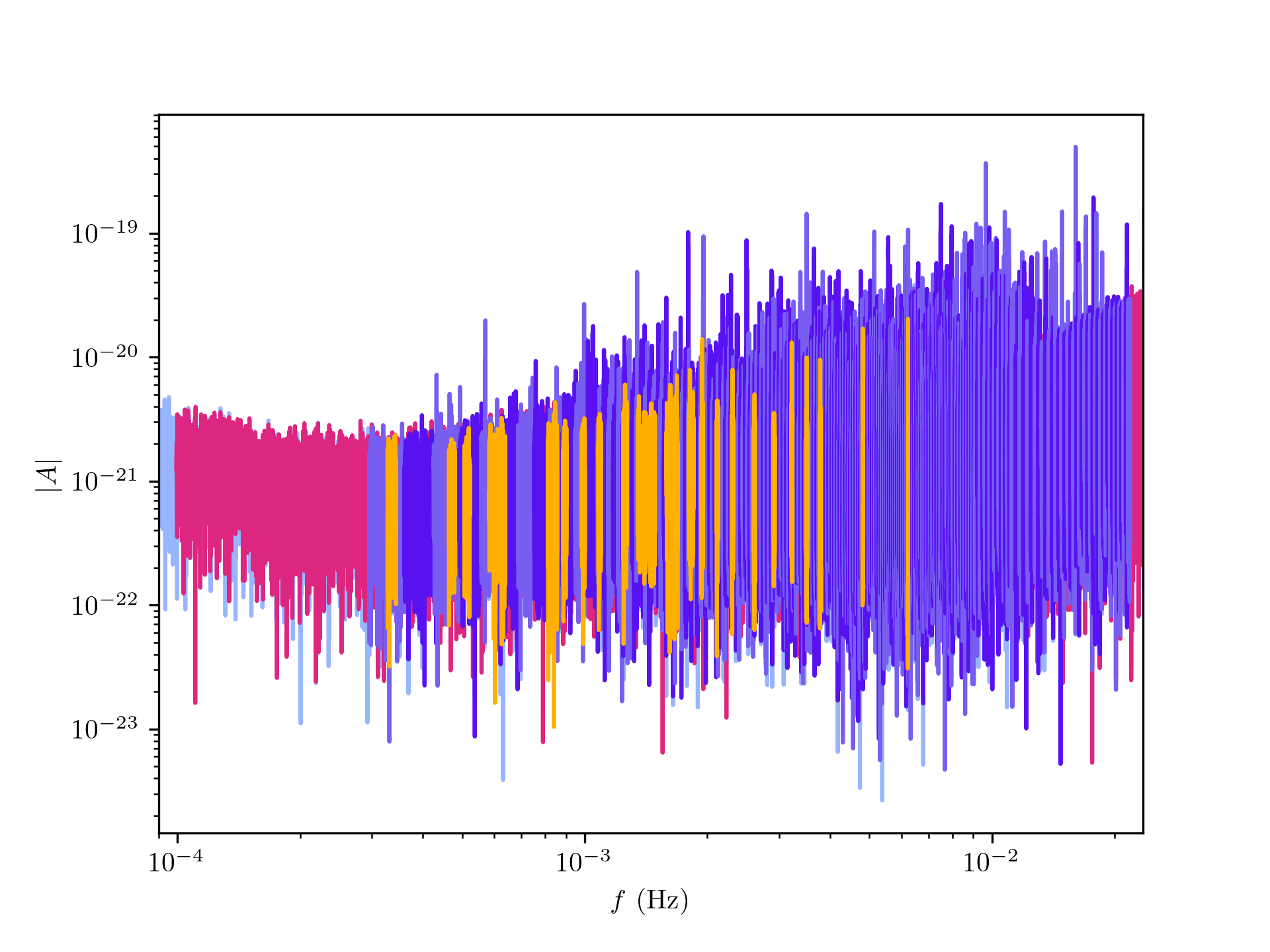}
    \caption{\small Amplitude spectral density (ASD) of TDI A channel data analyzed by each block of the sampler. The UCB segments are shown in alternating shades of purple. Note the changing bandwidth of the UCB segments, which are larger where the source-density is lower. VGB segments are orange. The MBHB (magenta) and noise models (blue) cover the full analysis band, with the noise model extending to slightly higher and lower frequencies for margin. This example uses the first 6 months of the \Sangria\ data. }
    \label{fig:data_segments}
\end{figure}

Fig~\ref{fig:data_segments} shows the amplitude spectral density (ASD) of the TDI A channel after the first six months of the \Sangria\ data. 
Through the remainder of the paper, the A channel will be used to visualize the data and/or signal models.
The differences between the A and E channel data are subtle and not informative at this level, though they are crucial for the analysis to decompose the observed signal into the the two GW polarization states.
The data shown in the figure are colored over the intervals being analyzed by different model components. 
The noise model (light blue) covers the full frequency range.  
The MBHB model (magenta) spans a similar bandwidth, though there is additional padding for the noise model to ensure that it extends beyond where the MBHB signals are in band during the time they are observable.
The UCB segments are shown in alternating light and dark purple bands. Though it is difficult to discern from the figure, especially due to the frequency axis being on a log scale, the width of the UCB segments is frequency-dependent, roughly tuned to use narrower segments where the source density is higher.
Finally the locations of each narrow-band segment for the targeted VGB analyses are shown in orange.
Throughout the remainder of the paper the same color scheme will be used to identify different model components:  Blue for noise, purple for UCBs, orange for VGBs, and magenta for MBHBs.

\begin{figure}[htp]
    \centering
    \includegraphics[width=0.45\textwidth]{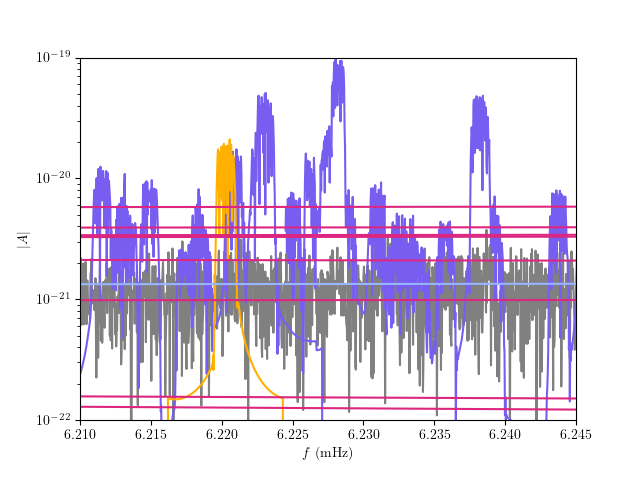}
    \caption{\small Same as Fig.~\ref{fig:reconstructions} but focused on a narrow frequency band near 6 mHz. The known binary in this segment of data (orange) is representative of how HM Cnc will appear in the LISA data.}
    \label{fig:reconstructions_zoom}
\end{figure}

For the headline demonstration of \TheGlobalFit\ at work, Fig.~\ref{fig:reconstructions_zoom} shows the reconstructed components of each part of the data model. 
The figure is showing the same content as Fig.~\ref{fig:reconstructions} but zoomed in to a narrow interval around 6 mHz containing one of the loudest currently known sources, HM Cnc~\cite{Roelofs_2010}, shown in orange. 
Here we can see all of the model components on display, with a densely-packed collection of UCBs in addition to HM Cnc all overlapping one another (purple), and the MBHB mergers sweeping through the band (magenta).  
The gray curve depicting the residual after all model components have been subtracted from the data is fit by noise model shown in blue.
Note that in this figure the uncertainty in the reconstructions is thinner than the line widths in the figure, as all of the sources in this interval have high signal to noise ratio (S/N).

\begin{figure}[htp]
    \centering
    \includegraphics[width=0.48\textwidth]{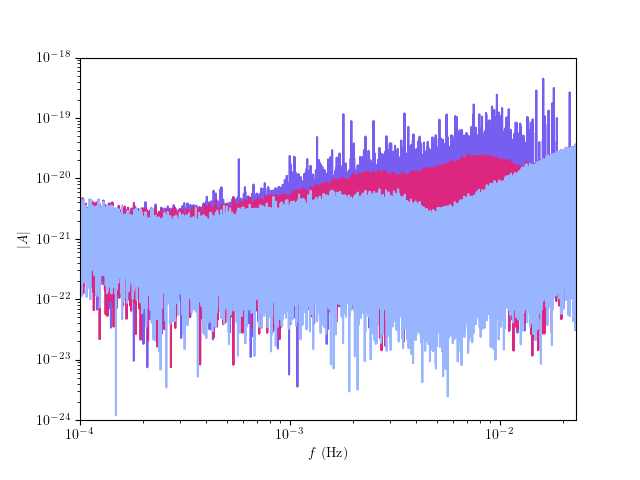}
    \caption{\small ASD of the data including all signals (purple), after removing the fit to the resolvable UCBs leaving behind only the MBHBs (magenta), and then the final residual after all signals in the fit are removed (light blue).}
    \label{fig:residuals_model}
\end{figure}

Broadening the aperture to the full analysis band of the demonstration, Fig.~\ref{fig:residuals_model} shows the original data's ASD which is dominated by the UCBs and is thus shown in purple.
Removing the resolved UCBs (and VGBs) leaves the magenta residual containing a bump in the spectrum from the combined signals of the MBHBs. 
The final (light blue) residual is after all of the resolved GW signals in the fit are subtracted from the data.
The remaining bump in the residual spanning ${\sim}3\times 10^{-4}$ to ${\sim}5 \times 10^{-3}$ Hz is due to the foreground of un-resolvable UCBs.

\begin{figure}[htp]
    \centering
    \includegraphics[width=0.45\textwidth]{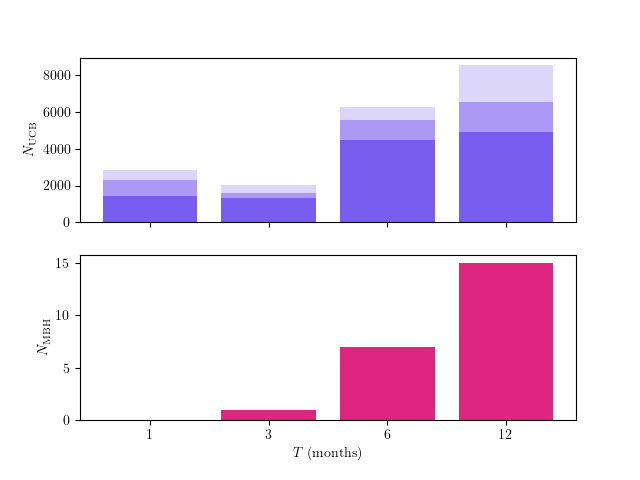}
    \caption{\small Number of UCB (top) and MBHB (bottom) detections as a function of observation time. The UCB detection number is the number of candidates from the maximum {\emph a posteriori} model after clustering samples by waveform match and then selecting candidates with $z>0.5$ (lightest shade), $z>0.9$ (medium shade), and those that uniquely correspond to a source in the injected population with a match of $m>0.9$. See Sec.~\ref{sec:demo.ucb} for a full explanation of the match.}
    \label{fig:detections}
\end{figure}

As described above the analysis is repeated on increasingly long epochs of the full data set, starting with the first 1.5 months of observations going up to the full year of data.
Analyses are conducted each time the data volume has doubled, resulting in analyses of 1.5, 3, 6, and 12 month segments.
As the observing time increases the number of detectable signals grows.  
For UCBs, which are continuous sources, this is due to the source building signal power over time and the improving frequency resolution of the data.
The MBHBs are transient sources so longer observation times provide more opportunity to catch a black hole merger in the act.
Fig.~\ref{fig:detections} shows the number of candidate detections in the source catalogs for the UCB (purple, left) and MBHB (magenta, right) models as the observing time increases.
The drop in the total number of UCB detection candidates between 1.5 and 3 months is unexpected.
However, the number of confident detections that are clearly associated with an injected signal increases, as will be described in section~\ref{sec:demo.ucb}.
Time-dependent analyses of data with such short observing times will be particularly sensitive to the initial orientation of the spacecraft constellation relative to the galactic center, and the modeling of the time-varying noise due to the galactic foreground~\cite{Digman_2022}.
While needing further study, our assessment is that the initial conditions of the LISA orbits and our admittedly incorrect assumption of stationary noise lead to this counter-intuitive result.

Having summarized the \TheGlobalFit\ performance on a data-wide level, we now take a detailed look at the performance of the individual model components by studying the properties of the recovered source catalogs and comparing them to the input populations.

\subsection{Noise Model}
The instrument noise model parameters used in \TheGlobalFit\ are not physically meaningful and so the primary diagnostics for the performance of the sampler are functional tests.
In each cycle of the \TheGlobalFit\ blocked MH sampler, the PSD model is fitting the residual after the current state of the UCB, VGB, and MBHB models have been subtracted from the data. 
That residual includes the instrument noise as well as the unresolved galactic foreground, referred to in the LISA literature as ``confusion noise,'' which is expected to be the dominant source of residual power between ${\sim}10^{-1}$ and ${\sim}3$ mHz. 
Exactly where the galactic foreground drops below the instrument noise depends on details of the galactic population of compact binaries~\cite{refId0}, the performance of the LISA instrument, and the observation time~\cite{Cornish_2017,PhysRevD.104.043019}.

\begin{figure}[htp]
    \centering
    \includegraphics[width=0.45\textwidth]{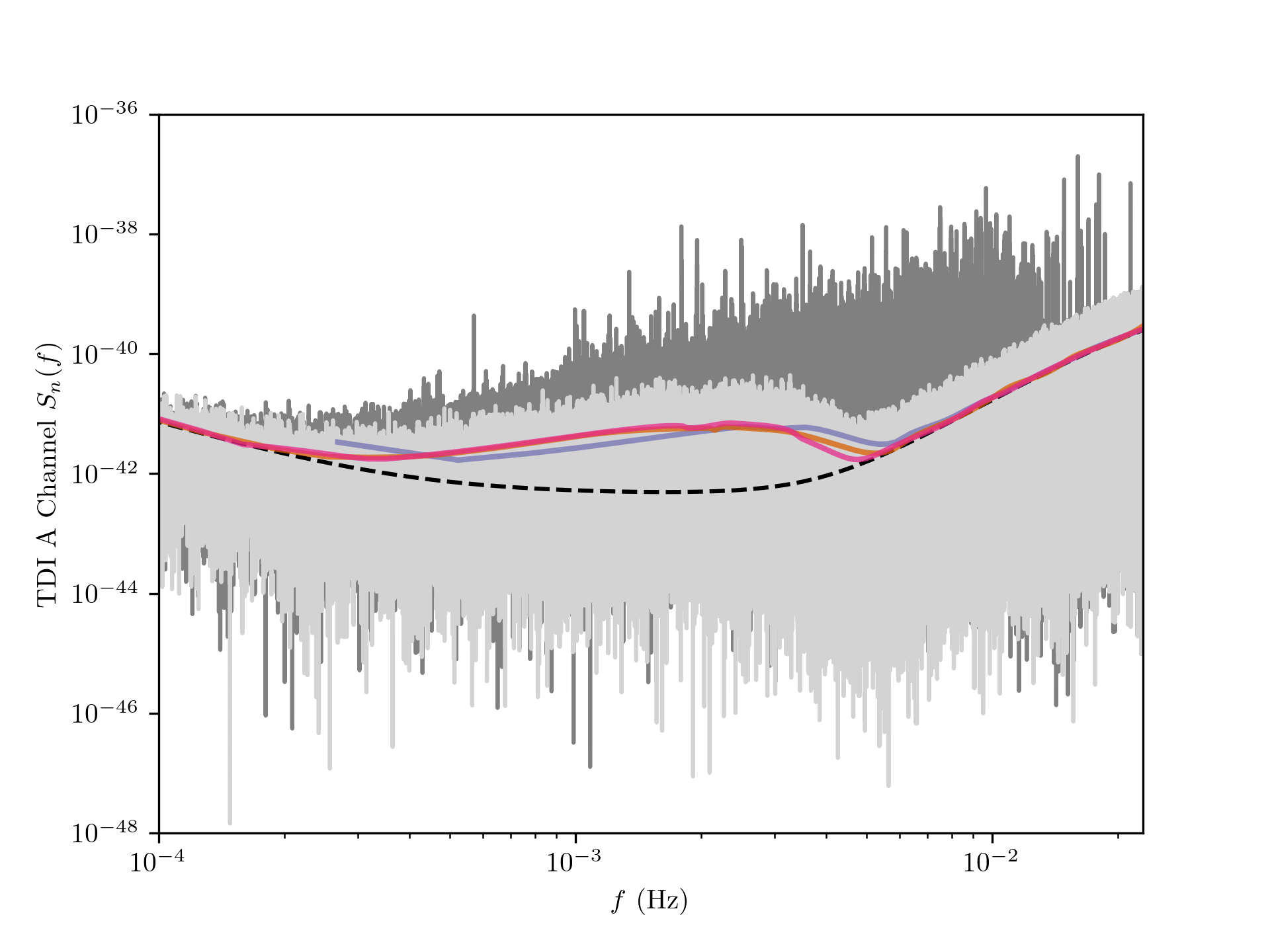}
    \caption{\small Median inferred noise PSD for three (purple), six (green), and 12e (orange) month observing times.  The black dashed line is the true PSD used when simulating the data. For reference, the dark gray is the power spectrum of the 12 month data, and the darker gray is the residual after removal of the UCB, VGB, and MBHB models.  The difference between the inferred and true PSD between $\sim 2\times 10^{-4}$ and $~\sim 6\times 10^{-3}\ {\rm Hz}$ is due to the unresolved galactic foreground, or ``confusion noise.''}
    \label{fig:psds}
\end{figure}

Fig.~\ref{fig:psds} shows the power spectrum of the 12 month data A channel (dark gray), and the residual after removal of a fair draw from the joint UCB+VGB+MBHB model (light gray).
The colored lines are the PSD fits from the noise model for the 1.5, 6, and 12 month runs.  
The 3 month result is omitted for clarity.
The black dashed line is the PSD used for simulating the instrument noise.
In each of the PSD fits the spline model used ${\sim}15$ to ${\sim}30$ control points.
The results clearly show how the prominent bump in the spectrum where the astrophysical foreground dominates initially grows with time across the band as the joint S/N of the galaxy increases, and then is slowly reduced as the UCB model is able to resolve more binaries, particularly at higher frequency. 
Outside of the interval dominated by the astrophysical foreground, the modeled PSD matches the simulated levels.
The 1.5 month PSD fit is truncated at low frequency because the bandwidth of the noise fit is dynamically set based on the signal content of \TheGlobalFit\ which does not include any MBHB merger signals in the first month of the \Sangria\ data, alleviating the need to model the PSD below ${\sim}0.3$ mHz. 

\begin{figure}[htp]
    \centering
    \subfloat[1.5 mo]{\includegraphics[width=0.15\textwidth]{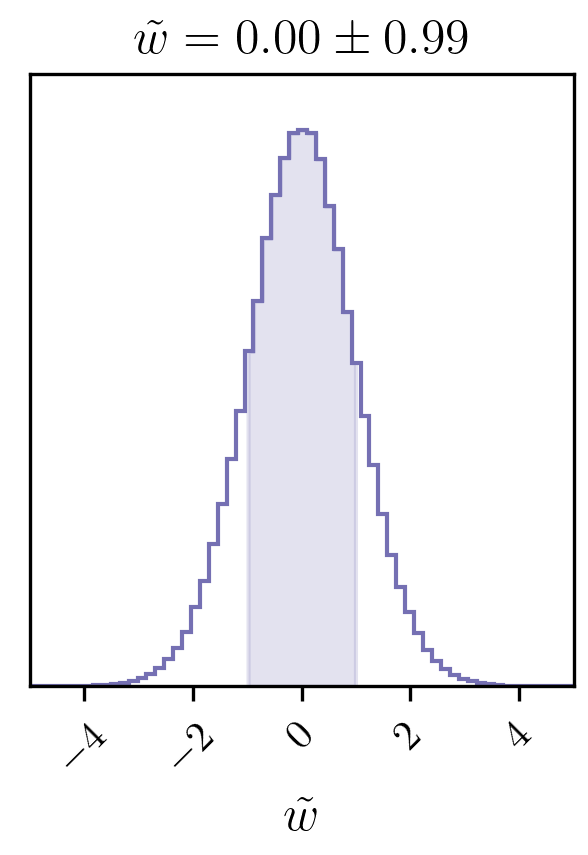}}\hfill
    \subfloat[6 mo]{\includegraphics[width=0.15\textwidth]{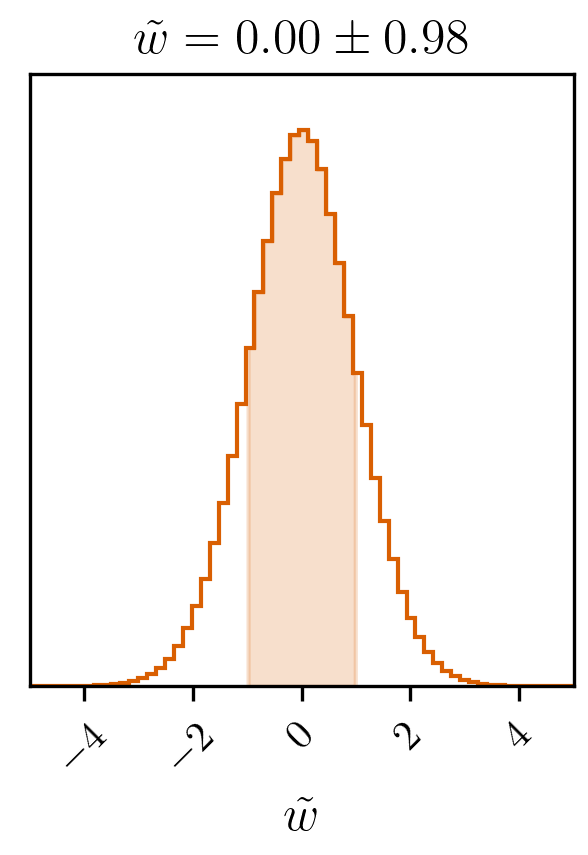}}\hfill
    \subfloat[12 mo]{\includegraphics[width=0.15\textwidth]{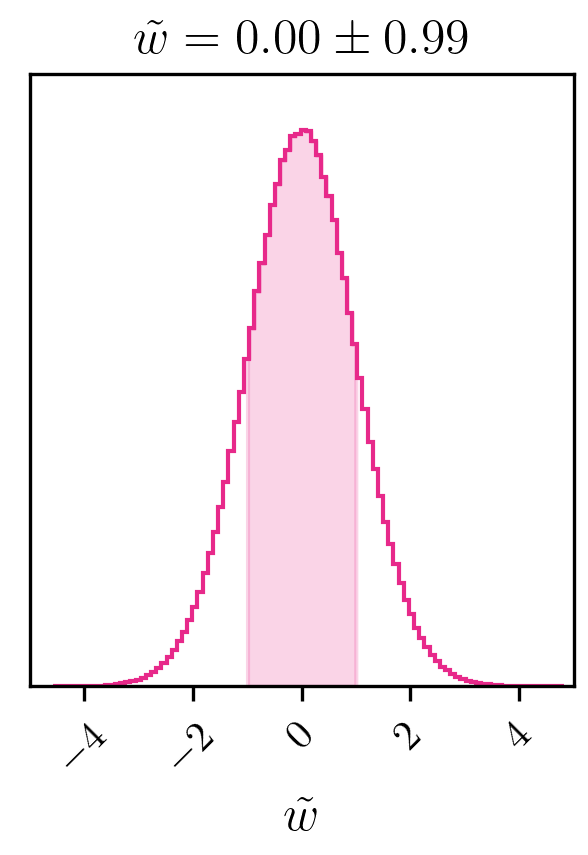}}\hfill
    \caption{\small Distribution of the whitened data $\tilde{w}(f) = \tilde{d}(f)/\sqrt{S_n(f)}$ for the 1 month [left], 6 month [middle] and 12 month [right] whitened residual data. If the PSD model is functioning correctly the whitened data should be distributed as a zero mean unit variance Gaussian. The mean and standard deviation computed from the whitened data are printed above each panel.}
    \label{fig:psd_whitening}
\end{figure}

A more quantitative assessment of the PSD model is possible by testing the whitened data $\tilde{w}(f)\equiv\tilde{d}(f)/\sqrt{S_n(f)}$ where the tilde denotes a Fourier transform, $d$ is the data, and $S_n(f)$ is the PSD.
The PSD is proportional to the frequency-dependant variance of the noise and therefore the whitened data should be consistent with a zero mean unit variance normal distribution $N[0,1]$.
Fig.~\ref{fig:psd_whitening} shows histograms of the combined real and imaginary components of the Fourier transformed whitened residuals for the 1.5 (left, purple), 6 (middle, orange) and 12 (right, pink) data. 
Displayed above each panel is the mean and standard deviation of the whitened residuals, in agreement with the expected results.
While the performance of the noise model passes the tests presented here we know that the model is incomplete and demands further development to meet the challenges of the real observing data.
The primary limitation of the current model is the implicit assumption that the noise is stationary.
In practice the LISA noise will have a time-varying PSD due to secular and random fluctuations of the instrument performance, as well as the cyclostationary modulations of the galactic foreground imparted by LISA's orbit~\cite{Edlund_2005,Digman_2022}. 
Generalizing the noise model using time-frequency methods~\cite{PhysRevD.102.124038} is an immediate priority for future development of \TheGlobalFit, and has ripple effects through the rest of the model components.

\subsection{UCB Catalog}\label{sec:demo.ucb}
\begin{figure}
    \centering
    \includegraphics[width=0.48\textwidth]{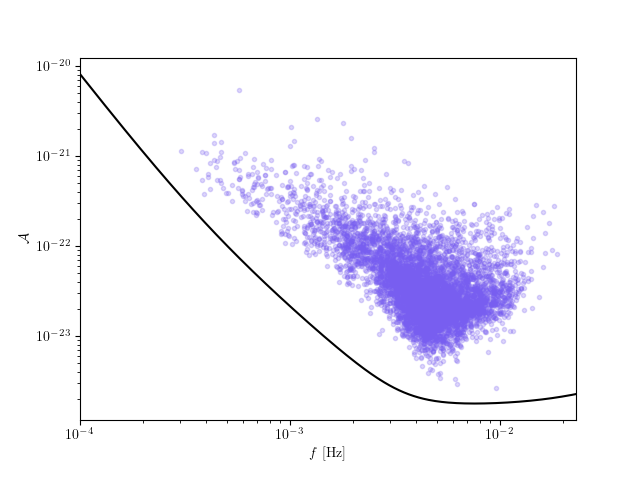}
    \caption{\small Scatter plot of frequency $f$ and GW amplitude $\mathcal{A}$ of UCBs in the 12 month source catalog. 
    The black line is an example LISA sensitivity curve.} 
    \label{fig:ucb_detections}
\end{figure}

The UCB catalog contains ${\sim}2000$ candidate detections after the first 1.5 months of observing, climbing to ${\sim}8500$ by the end of the 1 year \Sangria\ data set. 
Fig.~\ref{fig:ucb_detections} shows a scatter plot of the point-estimate frequency and GW amplitude parameters $(f,\mathcal{A})$ of the recovered sources in the catalog after analysis of the full 12 month \Sangria\ data.
The black line is a representative LISA sensitivity curve, so that the S/N of each source is proportional to its height above the curve. The ``cavity'' of sources above the curve at low frequency is due to the foreground of unresolved galactic binaries becoming the dominant noise source.

Distilling the output of the \gbmcmc\ sampler to a discrete list of catalog sources is a nuanced and lossy process, a detailed description of which is found in Ref.~\cite{PhysRevD.101.123021}.
To summarize: In each frequency segment the maximum a posteriori (MAP) number of source templates used to fit the data (i.e., the number of templates used most frequently in the RJMCMC sampler) is selected as the reference model.  
The posterior samples from that model are clustered into discrete catalog entries using the \emph{match} $m\equiv (h_i|h_j)/\sqrt{(h_i|h_i)(h_j|h_j)}$
between the waveforms computed from the chain samples at step $i$ and $j$ where $(\cdot|\cdot)$ is the standard noise-weighted inner product. 
The threshold for considering a chain sample as a member of a cluster is $m>0.8$.
The fraction of the total number of steps in the chain that have a sample in a particular entry is interpreted as a detection confidence $z$.  
The threshold for inclusion in the final catalog is $z>0.5$ i.e., that a catalog entry has a sample in more than half of the total number of chain steps in the MAP model. 
This is not a strict criteria, roughly equating to sources with a Bayesian odds ratio $>1$ being included in the model.
We therefore use an additional threshold of $z>0.9$ for catalog sources to be considered confident detections.
Neither the match requirement for inclusion as a sample belonging to a catalog entry, or the fraction of samples from the chain that an entry contains, are extensively tested or optimized through large scale injection studies.
Such critical work must be thoroughly undertaken in advance of using \TheGlobalFit\ or anything like it for production analyses.

\begin{figure}
    \centering
    \includegraphics[width=0.24\textwidth]{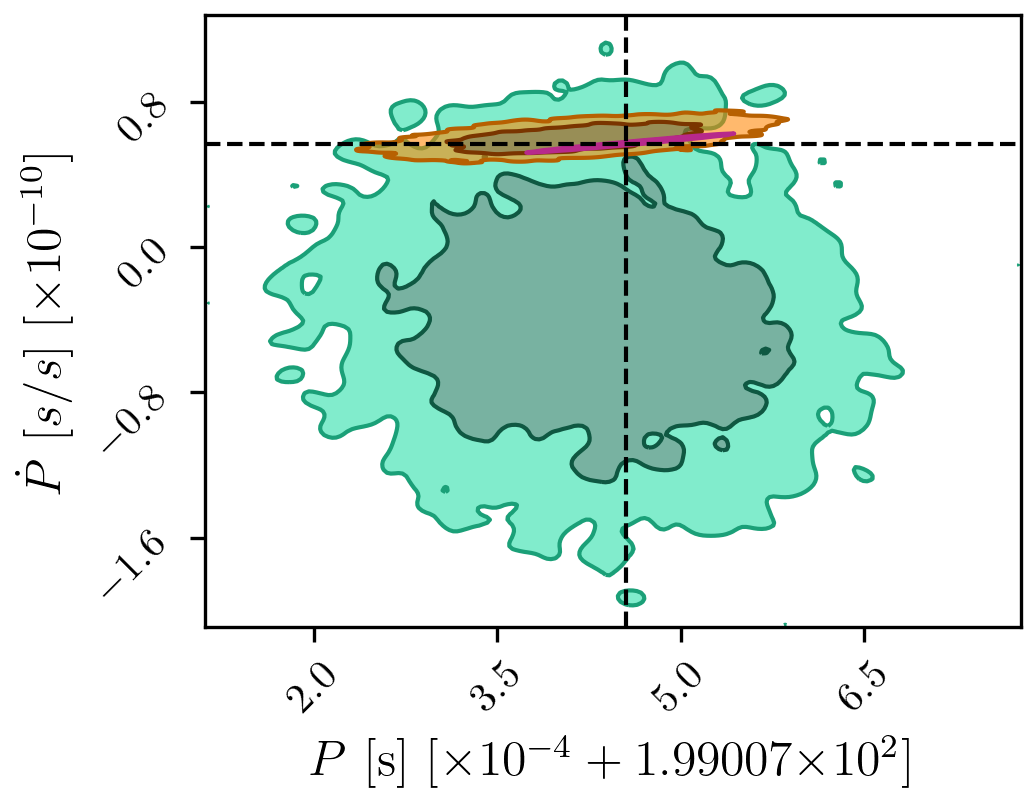}\hfill
    \includegraphics[width=0.24\textwidth]{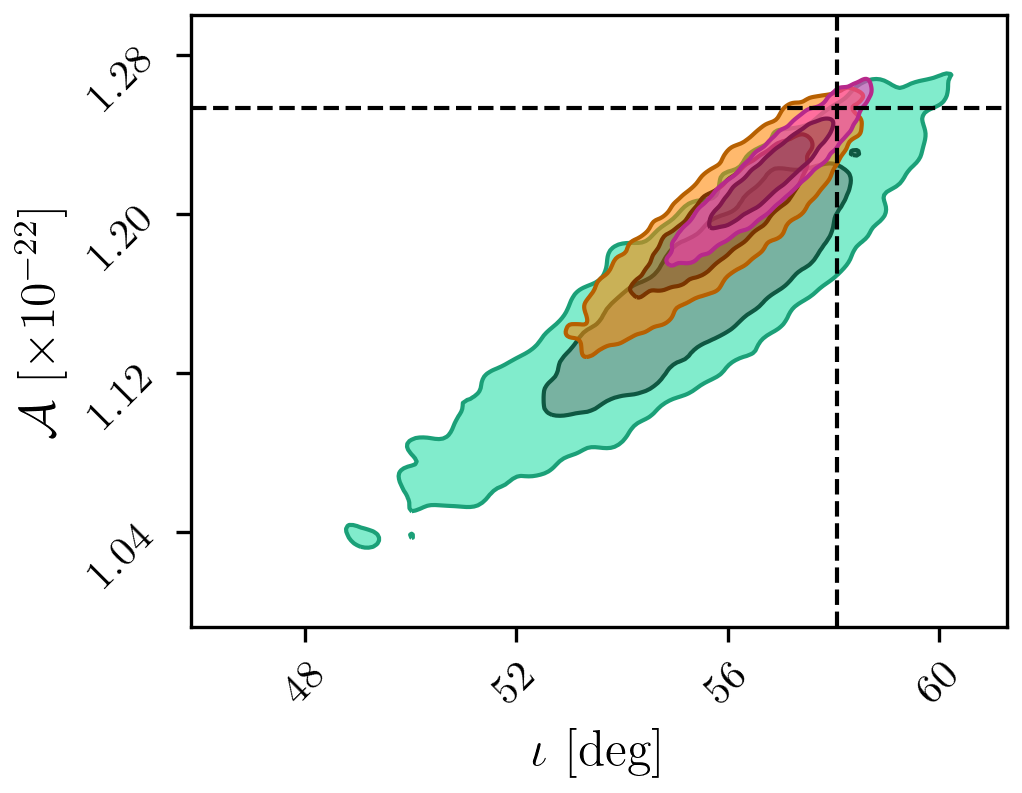}\hfill
    \includegraphics[width=0.24\textwidth]{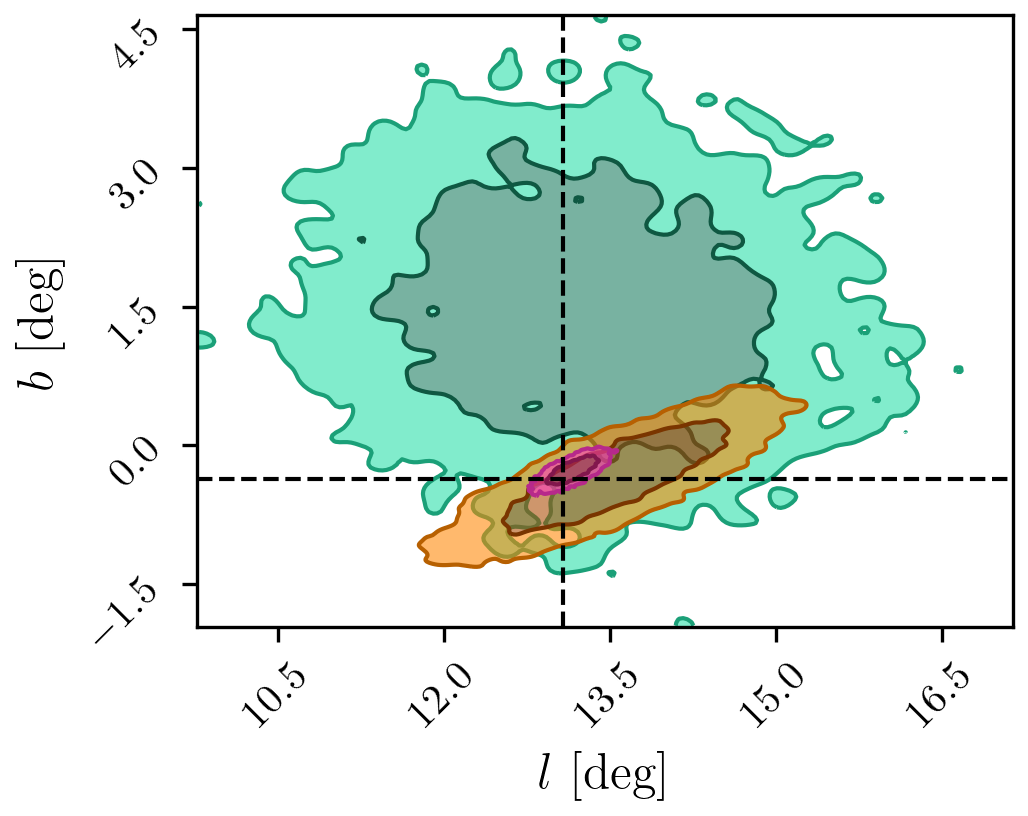}\hfill
    \includegraphics[width=0.24\textwidth]{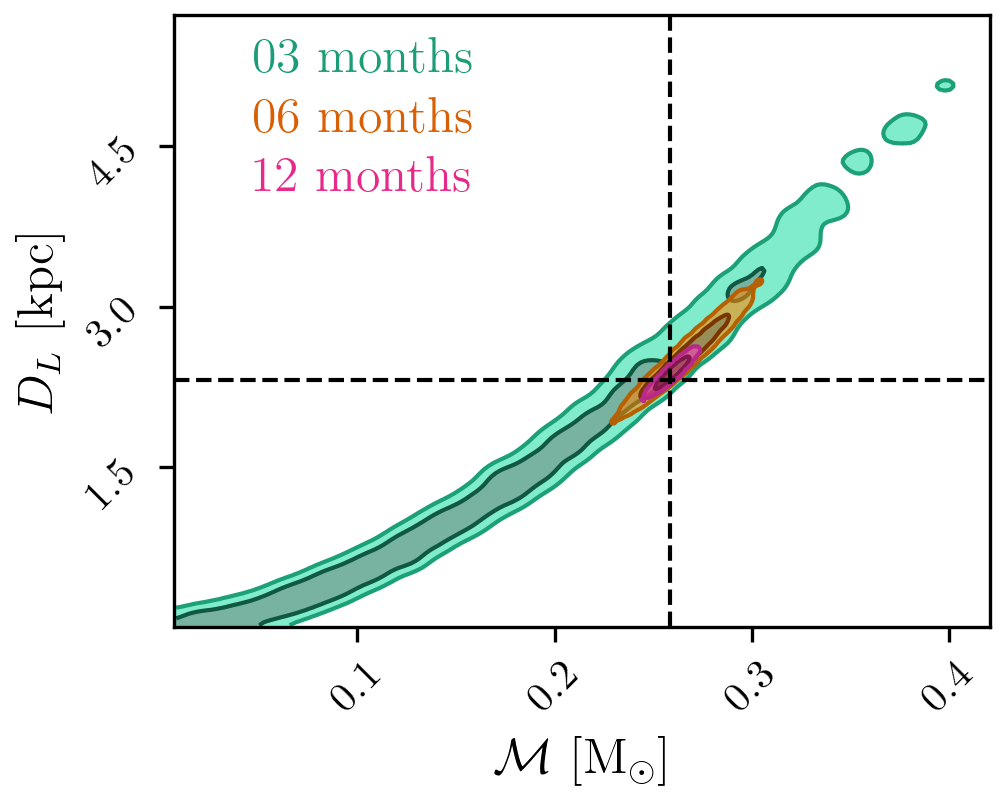}\hfill
    \caption{\small LDC0100498745 parameter estimation over time. Green, orange, and pink contours correspond to the measurement after 3, 6, and 12 months of observing with LISA.  The contours mark the 1 and 2$\sigma$ credible intervals. The sampling parameters from \gbmcmc\ are re-parameterized into orbital period and derivative $(P,\dot{P})$ [top left]; amplitude and inclination $({\mathcal A},\iota)$ [top right]; galactic latitude and longitude $(l,b)$ [bottom left]; chirp mass and luminosity distance $({\mathcal M},D_L)$ [bottom right].  The black dashed lines show the injected parameter values. }
    \label{fig:ucb_pe}
\end{figure}

As an example of the content contained for a single UCB, Fig.~\ref{fig:ucb_pe} shows a set of marginalized 2-dimensional posterior distributions for a high S/N binary found near $10$ mHz. 
UCBs in the \TheGlobalFit\ catalog are identified by their median frequency, so in the 12 month catalog this binary was labeled LDC0100498745.  
See~\cite{PhysRevD.101.123021} for a discussion and demonstration of how UCB candidates are traced through versions of the source catalogs from earlier analysis epochs (i.e. tracking how this particular source was labeled in the 6 month catalog, etc.).
Shaded contours are the 1 and 2$\sigma$ credible intervals, and the colors correspond to observing time with blue green, orange, and pink representing 3, 6, and 12 months respectively.
As with the color-coded source types, throughout the paper these colors (along with light-purple for 1 month) will consistently represent the observing times in subsequent figures.
The posteriors are represented in a different parameterization than is used in the sampling, to better match the observables customarily used by the EM observing community. 
The top left panel shows the orbital period $P$ and first derivative $\dot{P}$ of the binary system.
Note how the measurement precision of $\dot{P}$ increases more rapidly than other parameters.  
This is because the orbital evolution of the binary enters the phase as a $T^2$-dependent term, so the information accumulates more rapidly than the typical $\sqrt{T}$ scaling due to the increasing S/N of a continuous source.
The top right panel shows the gravitational wave amplitude and the binary's orbital inclination in degrees. 
The bottom left panel is the sky location in galactic coordinates, with $l$ as the galactic latitude and $b$ the galactic longitude. 
The bottom right panel are the chirp mass $\mathcal{M}$ and luminosity distance $D_L$ parameters derived from the GW observables assuming the orbital evolution is purely driven by emission of gravitational waves.  
The horizontal and vertical lines mark the parameter values for the injected signal, i.e. the ``right answer'' when we have the luxury of knowing the true source parameters.

\begin{figure*}
    \centering
    \includegraphics[width=0.49\textwidth]{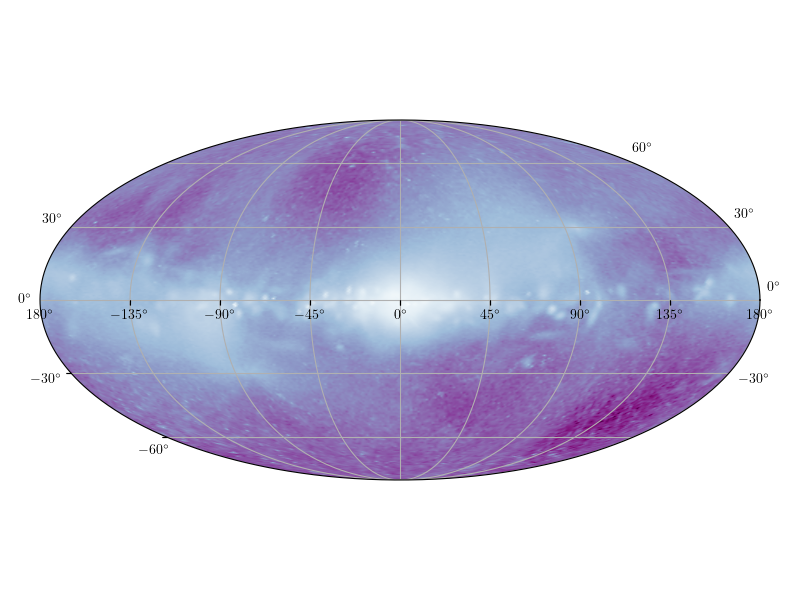}\hfill
    \includegraphics[width=0.49\textwidth]{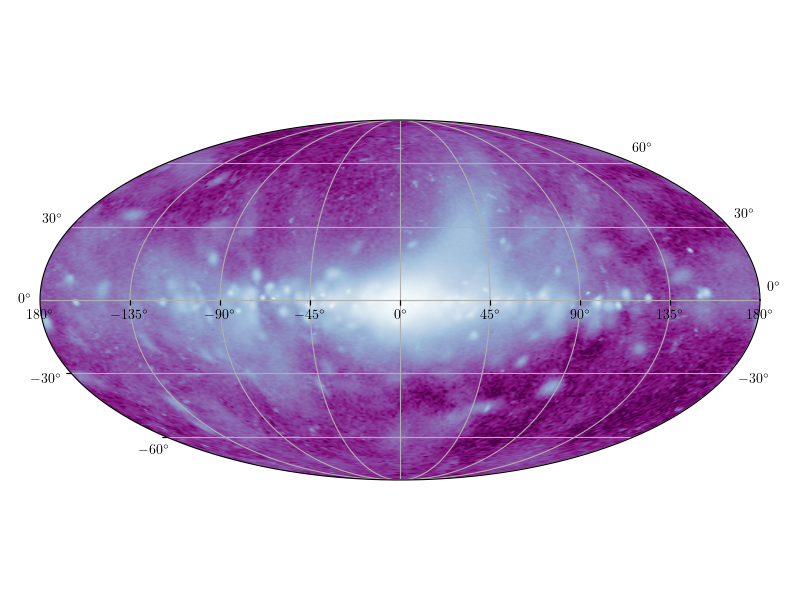}\hfill
    \vspace{-4em}
    \includegraphics[width=0.49\textwidth]{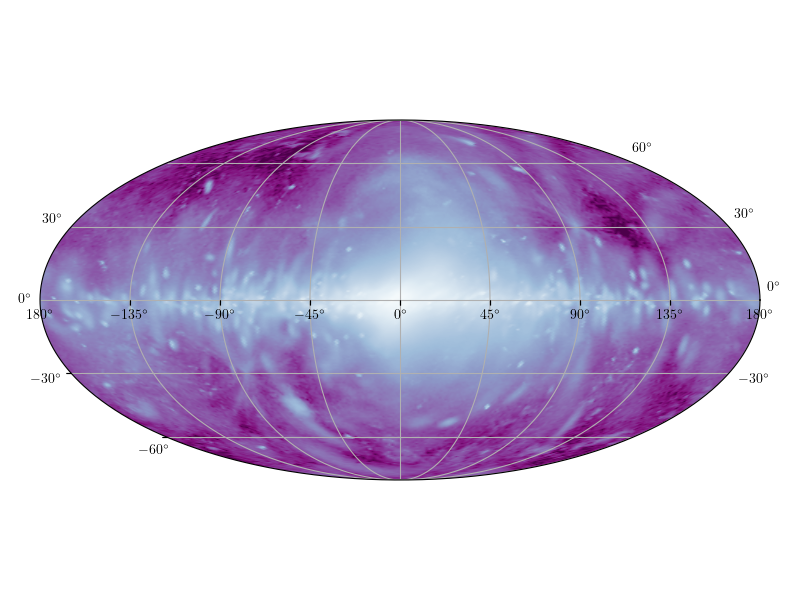}\hfill
    \includegraphics[width=0.49\textwidth]{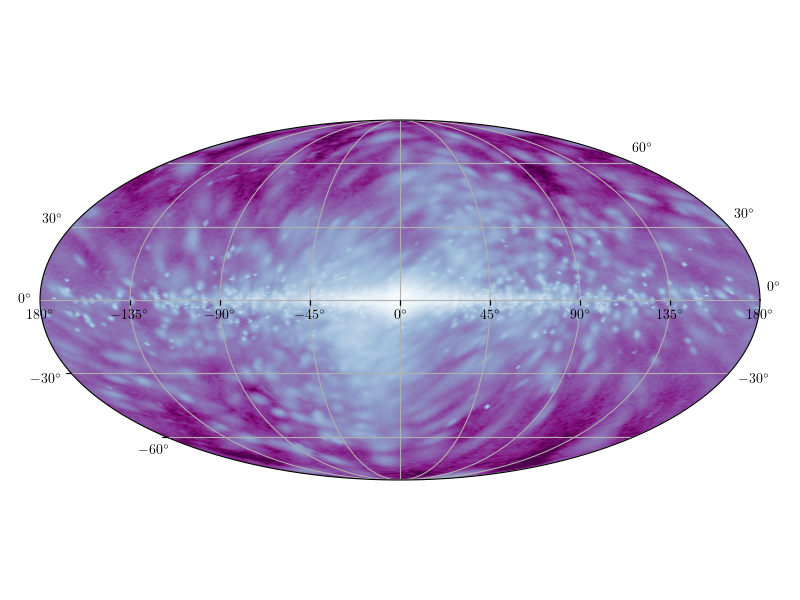}\hfill
    \caption{\small Maps of the source sky locations in galactic coordinates from the UCB catalogs after 1.5 [top left], 3 [top right], 6 [bottom left] and 12 [bottom right] months of observing, showing the increasingly clear reconstruction of the Milky Way disk and bulge structures.}
    \label{fig:ucb_sky}
\end{figure*}

One intriguing opportunity afforded by the LISA UCB catalog is to map the Milky Way's stellar remnant population.
Fig. ~\ref{fig:ucb_sky} shows the map of the UCB sky in galactic coordinates after 1.5, 3, 6, and 12 months from top left to bottom right.  
The maps are constructed by combining the posterior samples from all of the sources in the UCB catalog.
After only 1.5 months of observing large scale galactic features like the bulge and disk are evident in the maps.
The resolution of the image continues to improve as the observing time increases, revealing a remarkably clear view of the galactic disk and bulge with hundreds of well-localized sources.  
The quality of the image will steadily improve over the LISA mission life time, and will include distance information from the chirping binaries, enabling three-dimensional inferences on the spatial distribution of binaries throughout the galaxy~\cite{PhysRevD.86.124032}.

With the benefit of knowing the input source population, the observed UCB catalog is compared to the injected binaries to study the detection efficiency of the analysis. 
The primary metric for assessing the the quality of the inferred source catalog is the maximum match $m$ between the waveform computed from the point estimate source parameters in the catalog and the waveforms from the injected population.
For computational efficiency the match is only computed between the catalog waveform and an injected waveform if their frequency parameters are within 10 frequency bins of one another.

\begin{figure}[htp]
    \centering
    \includegraphics[width=0.45\textwidth]{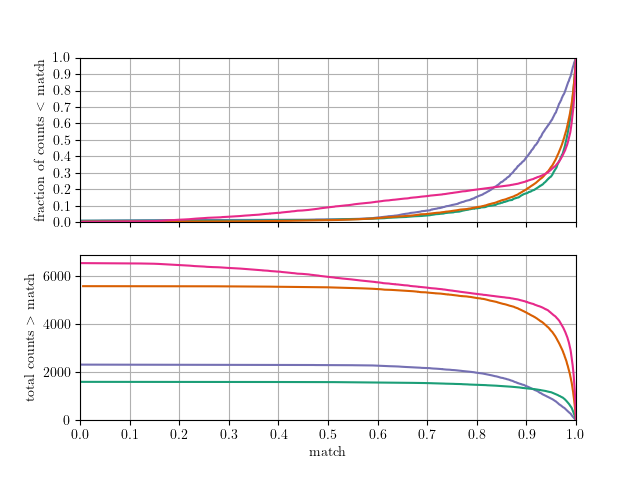}
    \caption{\small Top: Cumulative distribution function of matches.  Bottom: Un-normalized survival function of the match. Purple, green, orange, and pink curves are for 1.5, 3, 6, and 12 month observing times respectively.  Results are from confident ($z>0.9$) detections from the source catalog.}
    \label{fig:ucb_match}
\end{figure}

Fig.~\ref{fig:ucb_match} shows the distribution of the match values between the sources in the catalog and those in the input population.
For inclusion in the sample we select catalog candidates with detection confidence $z>0.9$ instead of the $z>0.5$ criteria for inclusion in the catalog.
The top panel shows the cumulative distribution function of the matches. 
The vertical axis is thus interpreted as the fraction of sources in the catalog with match below $m$. 
The bottom panel is the un-normalized survival function, i.e. the total number of sources in the catalog with match greater than $m$.
If we consider $m>0.9$ as a criteria for an unambiguous mapping between a source in the catalog and an injection then ${\sim}75\%$ (${\sim}80\%$) of the confidently identified  $(z>0.9)$ binaries in the 12 (6) month catalog exceed the criteria equating to ${\sim}5000$ (${\sim}4500$) sources.

The most striking feature in the results is a population of sources with matches below ${\sim}0.9$ that emerges in the 12 month catalog.
Previous analyses of simulated galaxies in LISA data have not shown a similarly-populated low match tail~\cite{Crowder:2006eu,PhysRevD.84.063009}, instead finding ${\gtrsim}90\%$ of sources with $m>0.9$. 
The comparatively high rate of low-match sources in the \TheGlobalFit\ UCB catalog demands further investigation and discussion.

\begin{figure*}[htp]
    \centering
    \includegraphics[width=\textwidth]{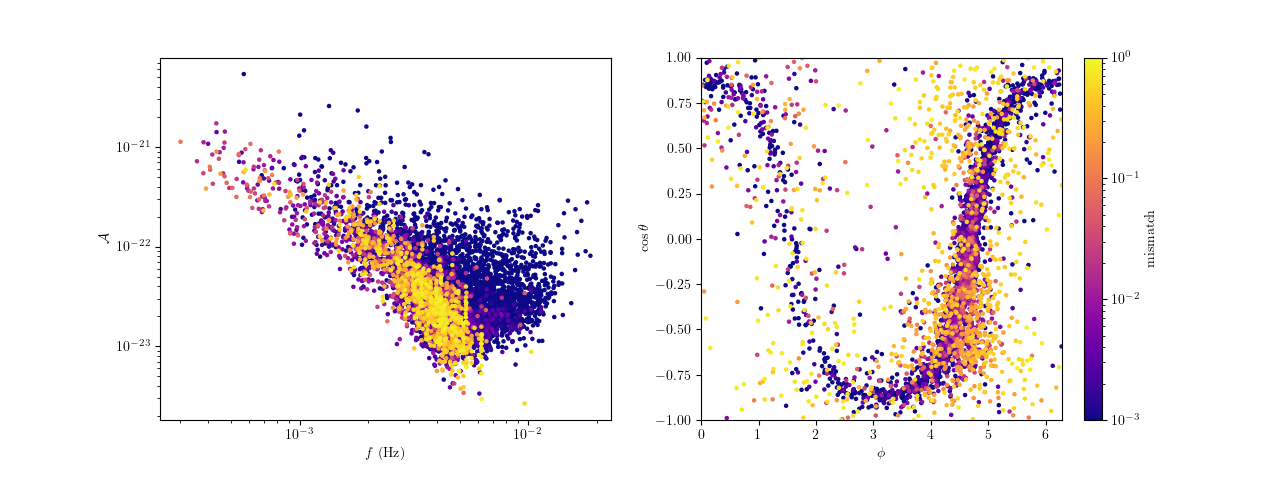}
    \caption{\small Scatter plot of point estimates for candidate sources in the UCB catalog after 12 months of observing colored by the minimum mismatch between the catalog waveform and the injected waveforms. Left: GW frequency-amplitude plane showing that most of the high mismatch sources are in the confusion noise regime below ${\sim}6$ mHz. Right: Sky location parameters in ecliptic coordinates, revealing that the high mismatch sources are preferentially located out of the galactic plane.}
    \label{fig:ucb_match_scatter}
\end{figure*}

To begin, compare the matches for each search over different projections of the full parameter space shown in Fig.~\ref{fig:ucb_match_scatter} displaying the frequency-amplitude plane on the left and the sky location on the right. 
Here the sky location is displayed using the sampling parameters (ecliptic coordinates) rather than galactic coordinates as in Fig~\ref{fig:ucb_sky}.
The source model in \gbmcmc\ is parameterized using ecliptic coordinates to minimize covariances between parameters, and thereby make it easier to sample the posterior.
In this figure each point in the scatter plot is colored by the waveform \emph{mismatch} defined as $1-m$, and the color map uses a log scale.
Thus cool colors are good matches, and warm colors are poor matches.

The majority of low match sources are found in the frequency interval between 1 and 6 mHz where the unresolved UCBs are the dominant source of noise.
One significant difference between the \TheGlobalFit\ analysis and previous UCB searches is the global spline noise model.
In references~\cite{Crowder:2006eu,PhysRevD.84.063009} each narrow-band frequency segment independently modeled the noise level effectively using a ${\mathcal O}(10^3)$ parameter piece-wise (and discontinuous) fit to the instrument noise.
The \TheGlobalFit\ noise model is constrained to be a continuous function, and is using model selection to determine the most parsimonious number of parameters, naturally making the noise model less flexible.
It holds together that the differences between a fixed (and high dimensional) piece-wise fit and the parsimonious spline model have a larger effect in the foreground-dominated part of the spectrum where there is stronger coupling between the UCB and noise models.

Another, and perhaps more impactful difference, is a difference in the prior used for the GW source parameters between the \Sangria\ analysis and previous demonstrations. 
Previous incarnations of the \gbmcmc\ sampler, and it's ancestors, have used a prior on the sky location parameters derived by assuming the sources followed the spatial distribution of the galaxy.
While that prior is still an option in \TheGlobalFit, it was intentionally not used for the \Sangria\ analysis in favor of a uniform prior on the sky location parameters. 
The choice to not use a galaxy prior was motivated by the idea of eventually performing a hierarchichal analysis where the posterior samples of the binaries in the catalog are used to constrain models for the spatial distribution of sources in the galaxy~\cite{PhysRevD.86.124032}.
Hierarchichal analyses are complicated by non-trivial priors on the posterior samples and so the choice was made to produce samples for the UCBs in the most accessible form possible.
The expectation was that this choice would produce larger uncertainties in the position reconstruction of individual sources, but an unintended consequence is the effect it had on the detection efficiency.
The right hand side of Fig.~\ref{fig:ucb_match_scatter} clearly shows sources with high mismatch are preferentially located outside of the galactic plane, whereas high match sources follow the expected ``U'' shape of the galaxy in ecliptic coordinates.

Mismatching sources in the catalog generally arise from two circumstances:  
Either a source template is fitting to blended contributions from multiple injections or multiple source templates are being used to fit a single injection.
The former (one-to-many) can be the most parsimonious solution (having the highest Bayesian evidence), or it could be due to modeling issues either from the noise or sky location prior while the latter (many-to-one) is clearly a problem with sampler convergence.

In the regime where a few templates are fitting a larger number of signals, clear attribution for why this was the preferred configuration of the model is difficult to assign.
However, it is true that from a strict model selection point of view if fewer templates can adequately fit features in the data, i.e. a larger combination of sources, the parsimonious solution is favored. 
Additional information, such as a prior that favors sources in the galactic plane, is needed to help further disentangle the overlapping sources.  
If the sky location prior and/or noise model were the predominant cause of the poor matches, why would it only appear in the 12 month analysis?
One possible explanation is that the recovered source catalogs from shorter integration times are going to be dominated by the loudest, most isolated, binaries in the population and as the observing time increases the search is able fit features in the data with lower intrinsic GW amplitude where the injected source density increases.
Additionally, the uniform prior on the sky location of binaries is worse for longer observation times. 
With short-duration observations the LISA UCB catalog will be dominated by near-by sources, especially at lower frequency, where a uniform distribution on the sky, while still not accurate, is closer to the observed population than at later times when the UCB catalog will sample the full galaxy.
The issues of source confusion and the dependence on priors for what constitutes a ``detection'' are many, nuanced, and require dedicated study. Another possibility is that the jump from 6 months to 12 months was too big a step for the Gaussian mixture model proposal which uses the posteriors from the shorter time span analysis as a proposal for the longer time span analysis.

\begin{figure*}[htp]
    \centering
    \subfloat{{\includegraphics[width=0.48\textwidth]{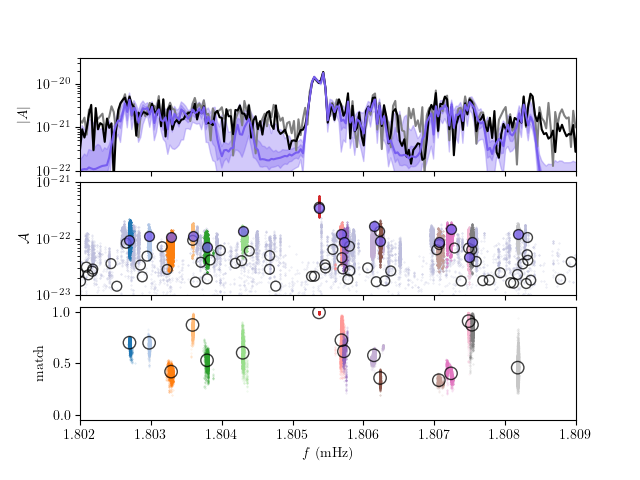}}}
    \subfloat{{\includegraphics[width=0.48\textwidth]{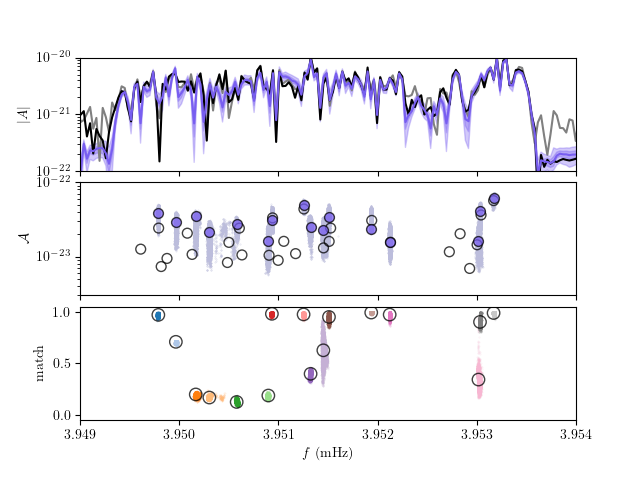}}}
    \caption{\small Investigation of possible causes for the low match population in the UCB catalog. Left and right panels focus on different frequency intervals towards the low (${\sim}1.8$ mHz) and high (${\sim}4$ mHz) end of the region where the false alarms were most frequent. The top panel shows ASD of the \Sangria\ data with MBHBs removed (gray), input UCB signal (black) and the joint posterior for reconstruction from the \gbmcmc\ sampler (purple).  The middle panels are a scatter plot of the UCB frequency-amplitude parameters for the injections (open circles), point estimate catalog entries (filled circles), chain samples (light purple dots), and posterior samples for the individual sources (multi-color dots). The bottom panel shows the maximum match between each posterior sample and an injected waveform (same colors as middle panel) and the match from the catalog point estimate used in the summary plots like fig.~\ref{fig:ucb_match}. The low-match sources are consistent with fitting combinations of injections.}
    \label{fig:ucb_match_investigation}
\end{figure*}

Fig.~\ref{fig:ucb_match_investigation} shows a pair of examples suspected of exhibiting the parsimony ``failure'' mode.
Both show results from a subset of a single UCB analysis segment from the 12 month data roughly bracketing the frequency range where the low match sources are most prevalent.
The top panels show the ASD of the simulated data after the MBHB signals have been removed (gray), the injected UCB model (black), and the posterior distribution for the recovered UCB model (purple).
The middle panels show the injected source parameters in the frequency-amplitude plane (open circles), the point estimate parameters from the UCB catalog entries (filled purple circles), unfiltered chain samples from \gbmcmc\ (light purple), and the posterior samples colored differently for each individual source in the catalog.
The bottom panel shows the match for each posterior sample in the same colors as the panel above, and open circles for the point estimate match used to construct the curves in Fig.~\ref{fig:ucb_match}.
In these examples the catalog entries with poor matches are typically found in parts of the frequency segment where there are numerous unresolved injections.
The posterior samples do not obviously favor one injected value over the other, yet the overall fit follows the injected signal (top panel).

\begin{figure}[htp]
    \centering
    \includegraphics[width=0.48\textwidth]{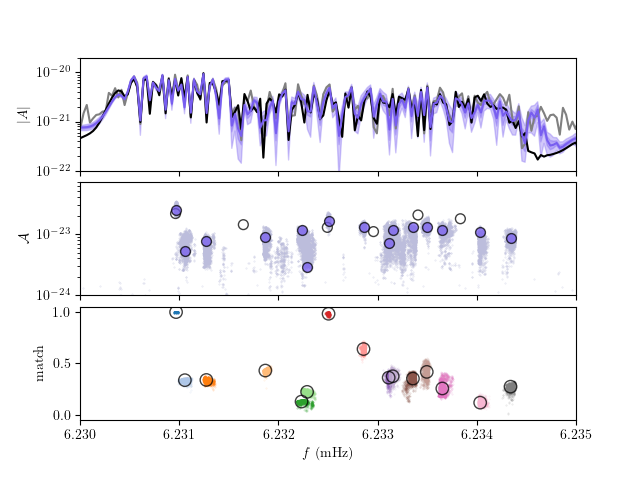}
    \caption{\small Same as fig.~\ref{fig:ucb_match_investigation} for an example where multiple templates were fitting a smaller number of injections, i.e. a clear example of a convergence failure for \gbmcmc.}
    \label{fig:ucb_match_investigation_overfit}
\end{figure}

An example of the overfitting problem, where multiple templates are fitting one source, is shown in fig.~\ref{fig:ucb_match_investigation_overfit} using the same format as ~\ref{fig:ucb_match_investigation}. 
In this segment there are six densely-packed injections, two of which are well-fit in the catalog.
The remaining four were fit by 13 templates in the 12 month analysis--a clear convergence error. 
Examples like figures~\ref{fig:ucb_match_investigation} and \ref{fig:ucb_match_investigation_overfit} will drive future development of the \gbmcmc\ pipeline.

To test our conjecture about the root cause of the high rate of low match sources in the catalog, we reanalyzed five UCB segments from the \Sangria\ data with the MBHB injections removed, and used the \gbmcmc\ sampler alone.
The frequency segments were chosen to be evenly spaced between 1 and 6 mHz to cover the regime where the low match catalog entries were most common.
Three different configurations were used to measure the effect of the different suspects for the low match population. 
The first uses proposals built from the 6 month \TheGlobalFit\ run, and the flat sky prior, just like the production \Sangria\ results. 
The only difference between the first example and the global analysis (ignoring covariances with the MBHB model) is the noise model, which in \gbmcmc\ is parameterized as a constant over the frequency interval of the segment.  
Assuming a constant PSD over the each analysis segment effectively amounts to a \emph{significantly} more flexible noise model compared to what is used in \TheGlobalFit.
The second configuration is the same as the first, but with proposals built from a 9 month \gbmcmc\ analysis of the segments to test whether the low match sources were due to convergence problems stemming from the time steps between analyses being too large, reducing the efficiency of the Gaussian mixture model proposals.
The final configuration reverts to the 6 month proposals but includes the galaxy prior as described in ~\cite{PhysRevD.101.123021}.

\begin{figure}[htp]
    \centering
    \includegraphics[width=0.45\textwidth]{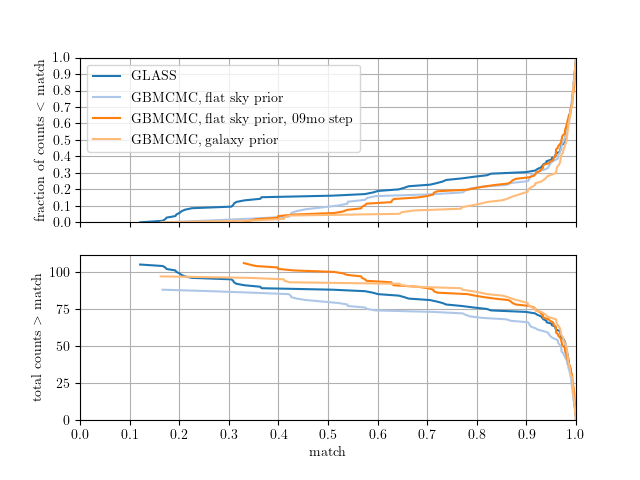}
    \caption{\small Same as fig.~\ref{fig:ucb_match} but for different test configurations of \gbmcmc\ on six frequency segments between 1 and 6 mHz to explore possible causes for the high fraction of low match sources in the \TheGlobalFit\ UCB catalog. The dark blue curves are the \TheGlobalFit\ results for the test segments. Light blue use the flat sky prior but a constant PSD model. Dark orange are the same configuration as light blue but include proposals generated from a 9 month run. Light orange uses the constant PSD model and a prior that prefers sources to be located in the galactic plane. Results are from confident ($z>0.9$) detections. In this limited test, using the galaxy prior improves both the catalog purity and the number of high-match sources in the catalog and shows that smaller time steps between global fit runs are advantageous.}
    \label{fig:ucb_match_tests}
\end{figure}

Fig.~\ref{fig:ucb_match_tests} shows the same type of results as fig.~\ref{fig:ucb_match} but only for the six test segments, and comparing the different test configurations with the global analysis.
Of the different configurations tested, using the constant PSD model slightly improves the purity of the catalog, but similarly reduces the total number of detections in the catalog. 
Including an intermediate 9 month analysis results in a higher number of detections and an improved match distribution, meaning that the overall convergence of the UCB model is heavily dependent on the efficiency of the proposals.
Including the galaxy prior has the largest influence on the results, both improving the catalog purity \emph{as well as} the total number of detections. 
While a full-scale study is needed to conclusively assess the performance of the different configurations, these results are suggestive that using a model for the galaxy in the analysis and shorter time steps between global fit processing will improve the quality of the UCB source catalog.
The sensitivity to the galaxy prior also reinforces the fact that modeling choices have a significant impact on the resulting inferences, especially near the detection threshold.

\subsection{VGB Catalog}
Analysis and interpretation of the VGB catalog is more straightforward due to the model using informative priors derived from EM observations. 
Because the sampler is assuming a single source at known orbital period and sky location, the most relevant parameters to be measured by LISA are the GW amplitude and binary inclination. 
In the event that the LISA observation time is not long enough for the source to be ``detectable,'' the posterior samples are useful for setting inclination-dependent upper limits on the amplitude, and therefore the combined chirp mass of, and distance to, the binary.

\begin{figure}
    \centering
    \subfloat[AM CVn]{{\includegraphics[width=0.22\textwidth]{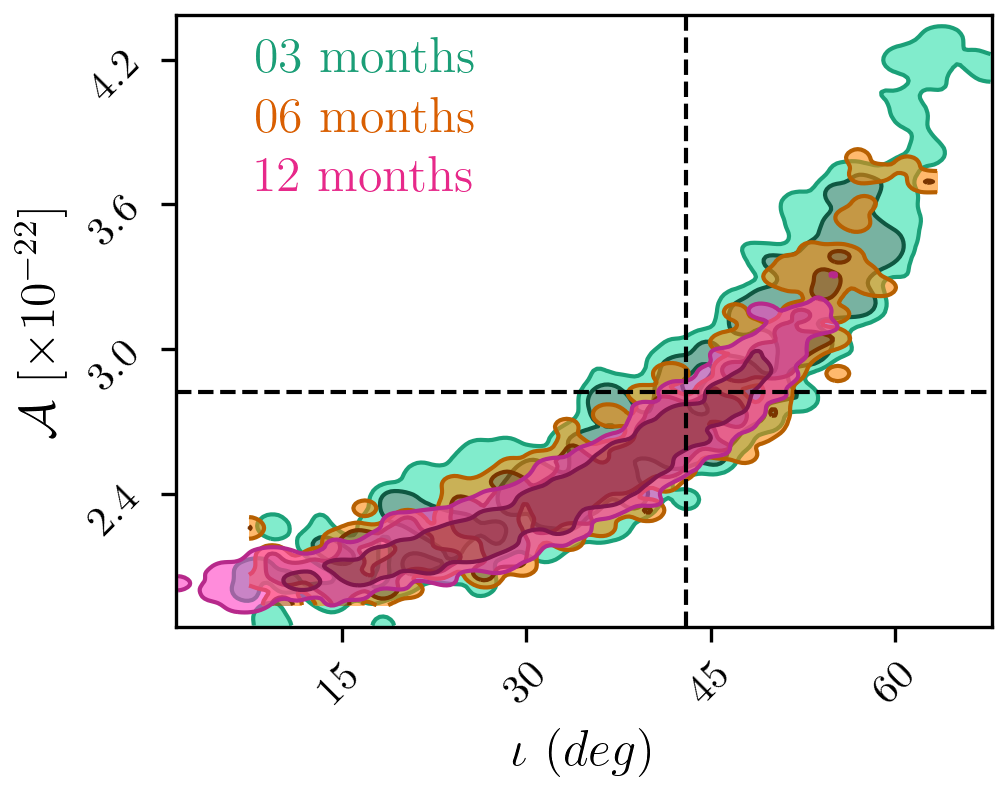} }}\hfill
    \subfloat[HM Cnc]{{\includegraphics[width=0.22\textwidth]{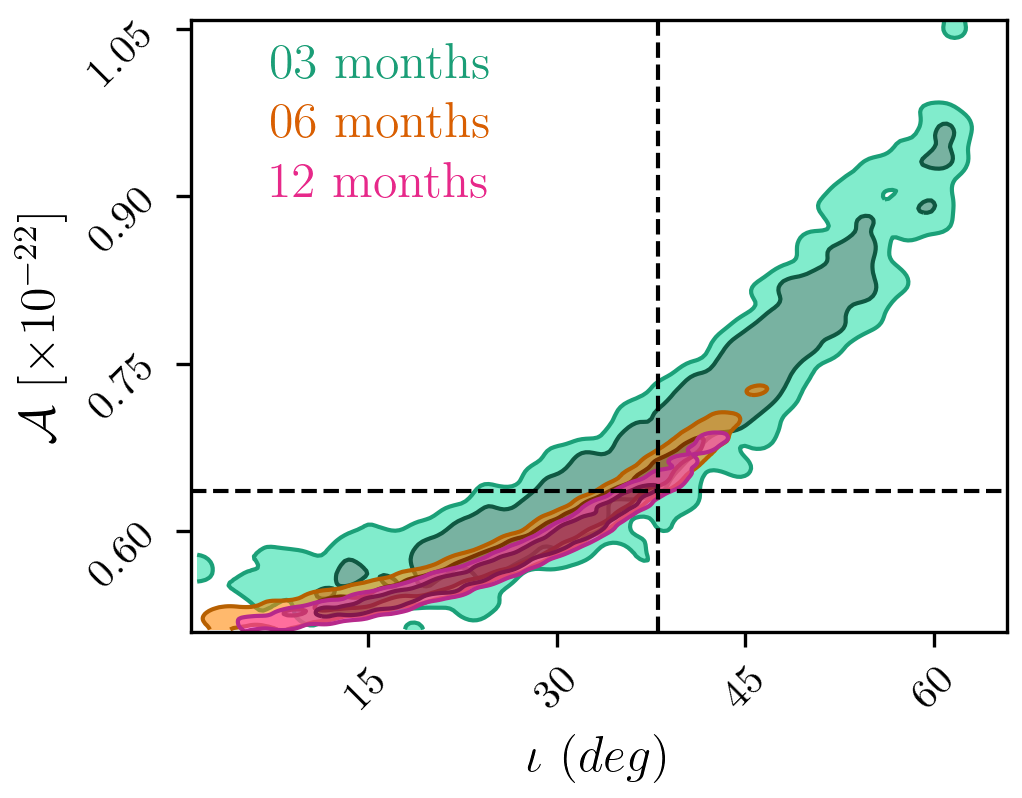} }}\hfill
    \subfloat[UCXB 4U1820-30]{{\includegraphics[width=0.22\textwidth]{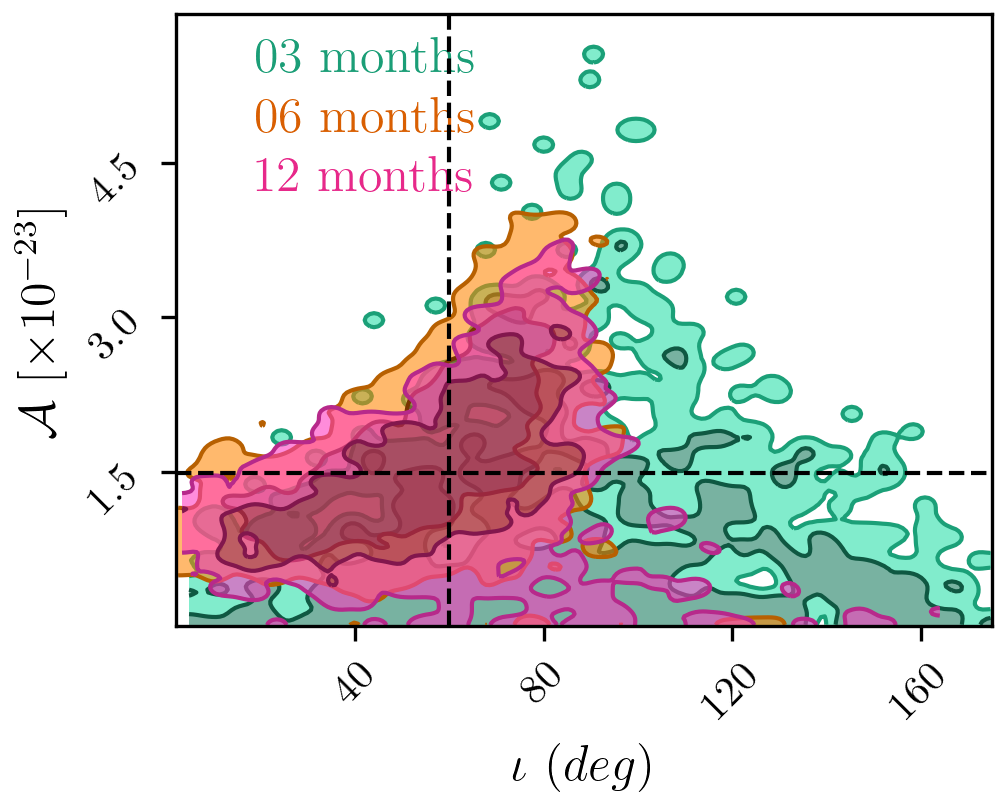} }}\hfill
    \subfloat[CX0GBS J1751]{{\includegraphics[width=0.22\textwidth]{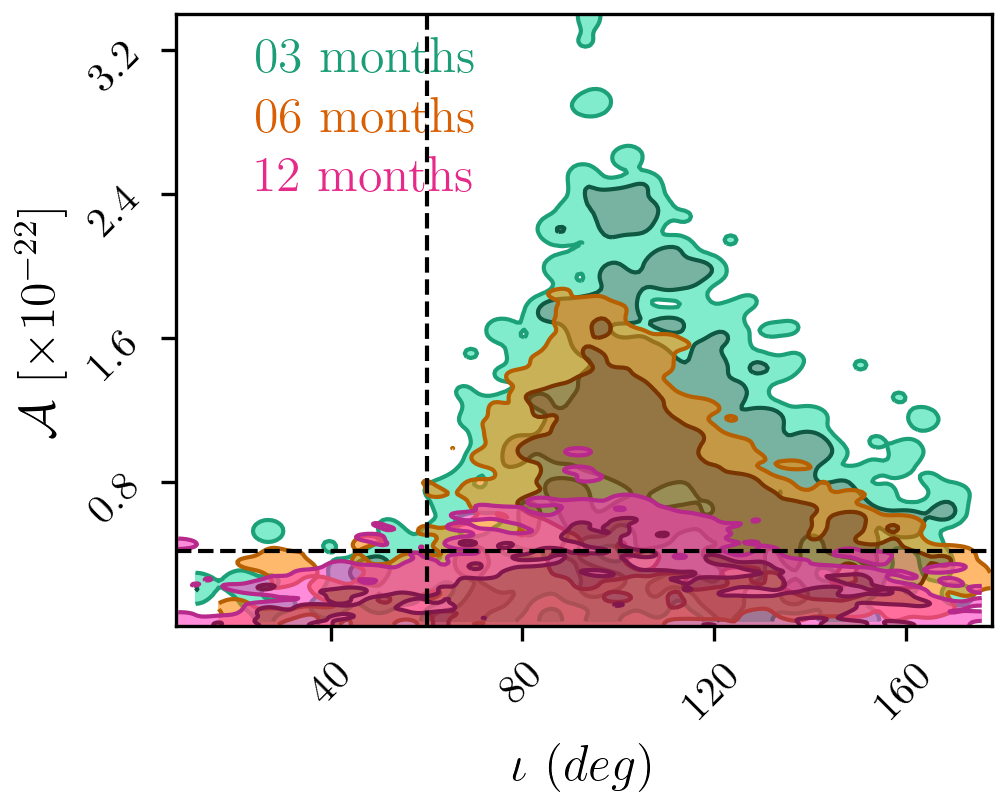} }}\hfill
    \caption{\small Variety of results from targeted analysis of known binaries in the \Sangria\ data showing measurement of the GW amplitude and binary inclination as a function of observing time, comparing 3 (green), 6 (orange), and 12 (magenta) month observations. (a) AM CVn is a straightforward example of a strong LISA source properly identified early in the observing campaign with inferences steadily improving over time. (b) HM Cnc shows similar behavior as AM CVn but at higher S/N. (c) UCXB 4U182030 shows how a binary will transition from being undetected, with the analysis providing upper limits on the amplitude (green) to a point where the binary parameters will be constrained (orange, magenta). (d) CX0GBS J1751 is an example of improving upper limits with observation time. This binary will require longer observing times to be constrained.}

    \label{fig:vgbs}
\end{figure}

Fig.~\ref{fig:vgbs} shows four representative examples from the full set of known binaries comparing inferences between 3, 6, and 12 month observations (green, orange, and magenta).  
The two dimensional posteriors present the 1 and 2$\sigma$ contours, and the dashed black lines mark the injected parameter values.
The top two panels show results for AM CVn and HM Cnc--two of the canonical known binaries that are identifiable early in the LISA observations.
The bottom left panel is for the ultra compact X-ray binary UCXB 4U1820-30 which transitions from a regime where upper limits are set after 3 months of observing, to a constraint in the 6 and 12 month catalogs, indicated by the open contours in green to the closed contours in orange and magenta.
Finally, source CX0GBS J1751 remains undetectable by \TheGlobalFit\ after 12 months of observing but note that the upper limit inferred for the amplitude decreases over time.

\subsection{MBHB Catalog}
\begin{figure}
    \centering
    \includegraphics[width=0.45\textwidth]{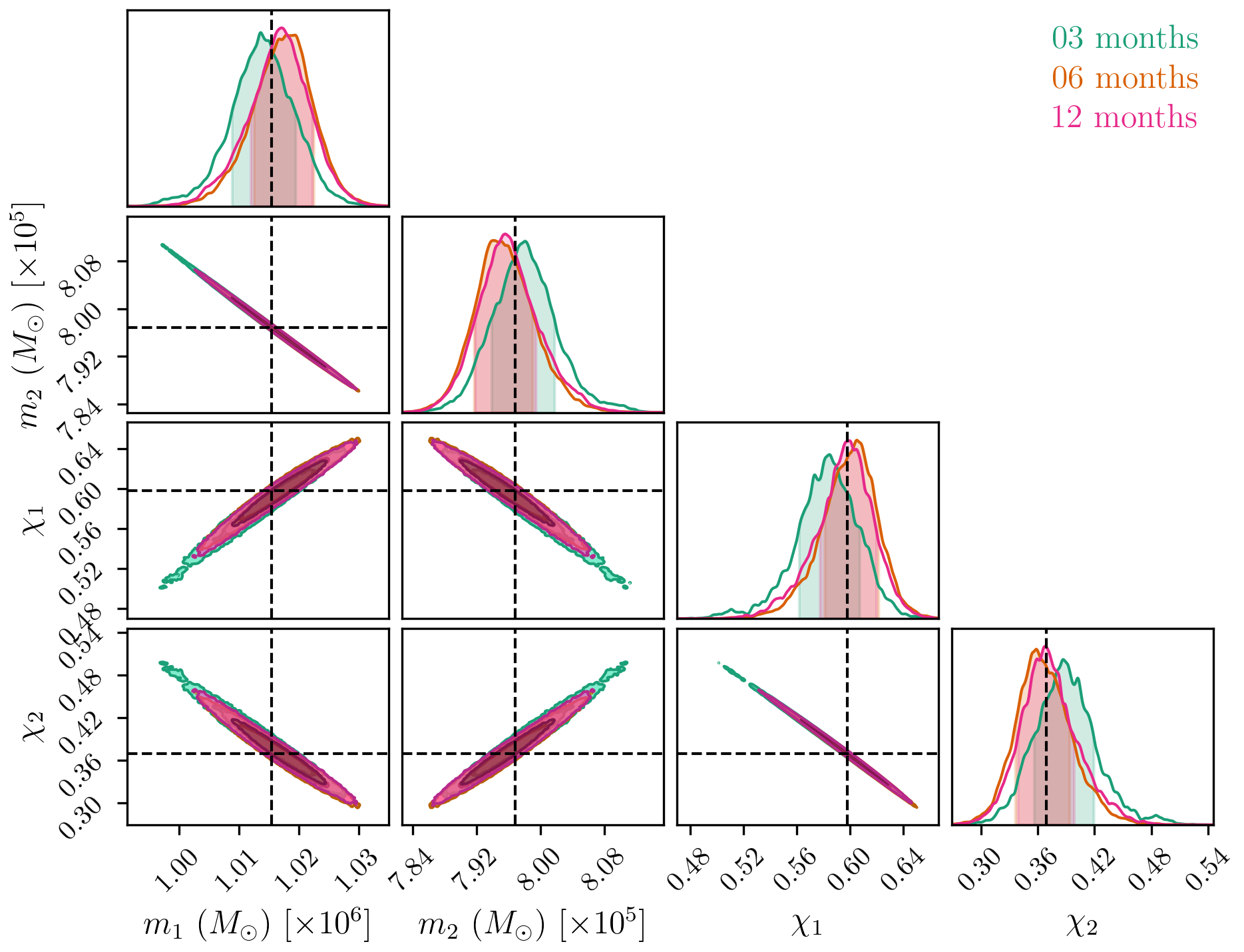} 
    \caption{\small Marginalized posterior distributions for the mass $m$ and dimensionless spin $\chi$ parameters of the first MBHB to merge in the \Sangria\ data, during the second month of the simulated data. The dashed lines mark the parameter values for the simulated signal. Note the parameter estimation improves after the signal has left the LISA band because the galactic foreground decreases as the UCB model resolves more binaries.}
    \label{fig:mbh_pe}
\end{figure}

The final parts of the \TheGlobalFit\ analysis to investigate are the MBHB results.
The most interesting single example is the first MBHB to appear in the \Sangria\ data, which merges during the second month of the simulated observations.
The simulated source also happens to be one of highest S/N binaries in the population and is observable in three of the different analysis epochs used for this demonstration of \TheGlobalFit.
Fig.~\ref{fig:mbh_pe} shows the posterior distribution function for the intrinsic source parameters: $m_1$ and $m_2$ are the masses of the black holes in the binary while $\chi_1$ and $\chi_2$ are their respective dimensionless spin parameters.
Recall that the data and MBHB model both currently assume the binaries have BH spin aligned with the orbital angular momentum vector--an assumption that is not valid in nature but made out of convenience at this stage of development for simulations and pipelines.  
As with other examples, the color indicates observing time and the dashed lines mark the injection values.

What is remarkable about fig.~\ref{fig:mbh_pe} are the changes in the posteriors over the observing time even though the binary merged in month 2 of the simulated data.
The subtle reduction in the width of the posteriors is due to the improved foreground subtraction from the UCB model, which effectively lowers the noise level, and thereby increases the S/N, of the MBHB mergers.
Because of the global nature of LISA analysis, inferences from transient sources will continue to improve long after they have left the measurement band.

\begin{figure*}
    \centering
    \subfloat{{\includegraphics[width=0.4\textwidth]{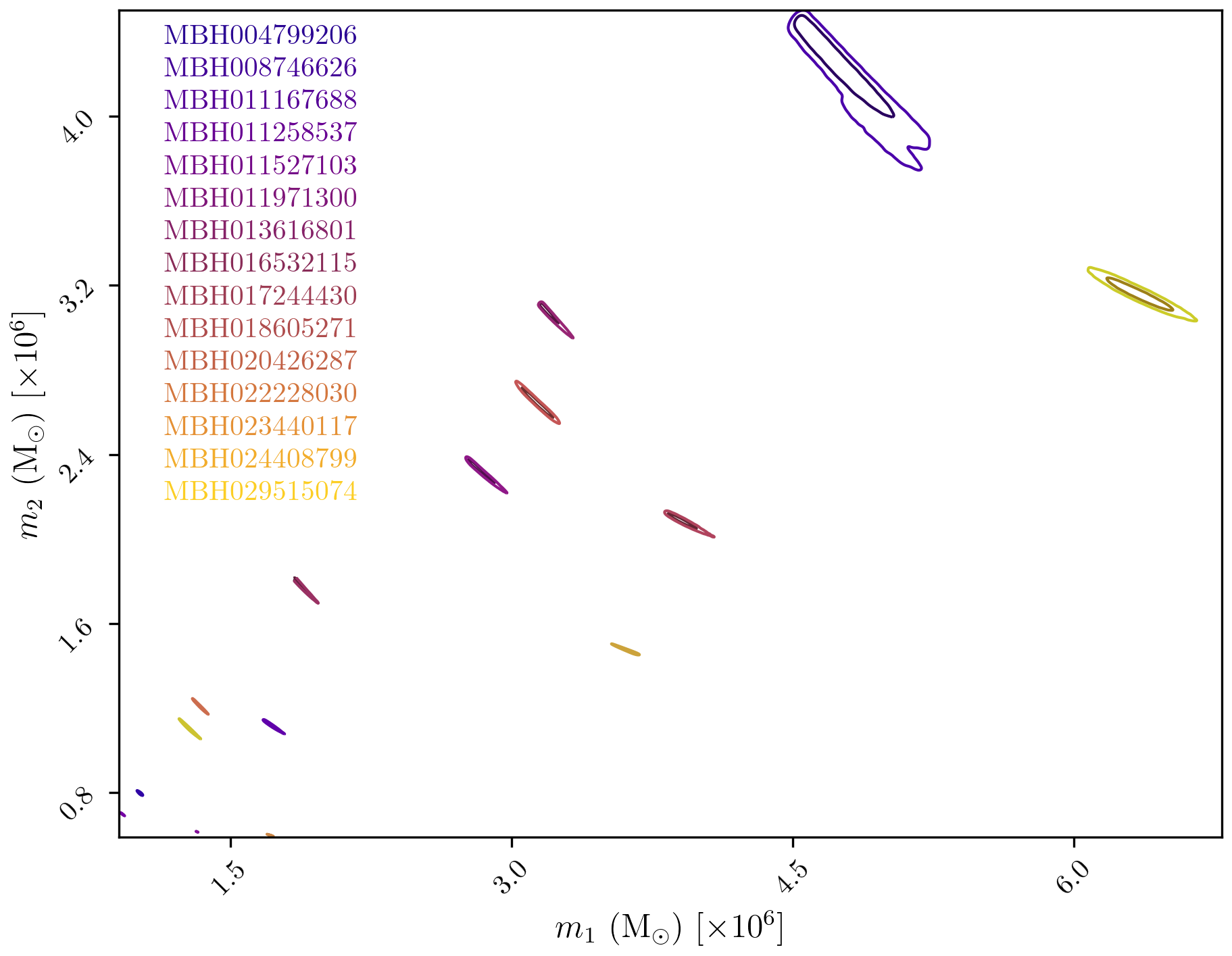} }}
    \qquad
    \subfloat{{\includegraphics[width=0.4\textwidth]{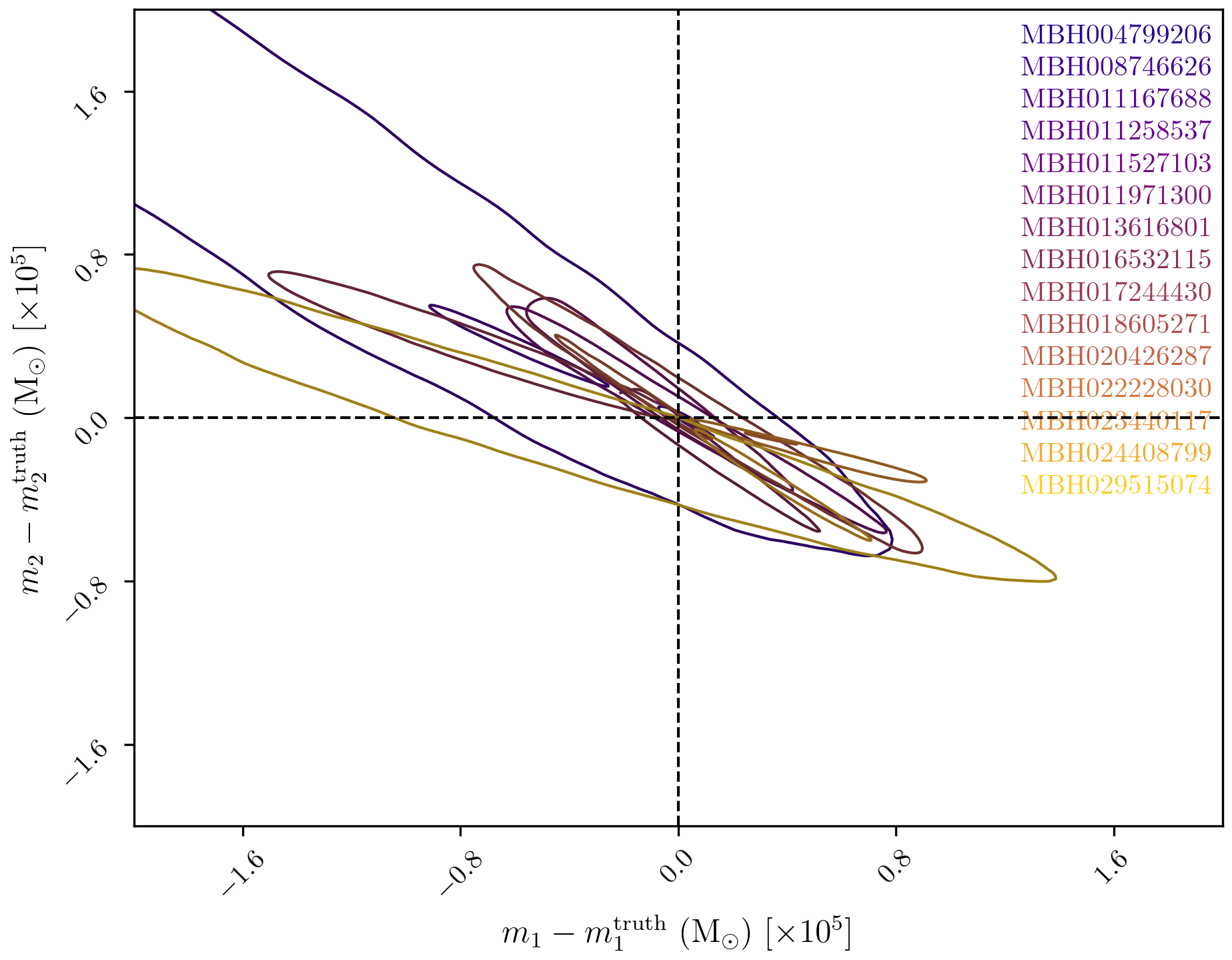} }}
    \qquad
    \caption{\small Mass posteriors for the entire observed MBHB population in the 12 month \Sangria\ data. Left: 1 and 2$\sigma$ contours for the inferred masses.  Right: 1$\sigma$ contours for the binaries shifted by the true values for each simulated source, such that (0,0) marks the injected parameter location.}
    \label{fig:mbh_mass_population}
\end{figure*}

Moving on to the full MBHB population, fig.~\ref{fig:mbh_mass_population} shows the posterior distribution function for the mass parameters from each of the 15 MBHBs injected into, and recovered from, the \Sangria\ data.
The MBHBs are labeled in the \TheGlobalFit\ catalog by the merger time (in seconds) relative to the start of observations.
The input population covers a wide range off masses and the posteriors are generally well-constrained due to the large number of in-spiral cycles combined with the strength of the merger signal.
To compare against the true values from the simulations, the right-hand panel shows the same posteriors but shifted by the injected mass values, so the point (0,0) marks the truth. 
For visibility, only the $1\sigma$ contours are shown on the right hand side. 
All but one of the binaries contain the truth value inside of the $2\sigma$ contour. 
The one source whose injected value is outside of the bulk of the \TheGlobalFit\ posterior is the lowest mass binary in the input population.
The same bias is seen in other prototype analyses and is suspected of being the results of artifacts in the \Sangria\ data from the waveform simulation process~\footnote{S. Babak, private communication}.

\begin{figure*}
    \includegraphics[width=0.31\textwidth]{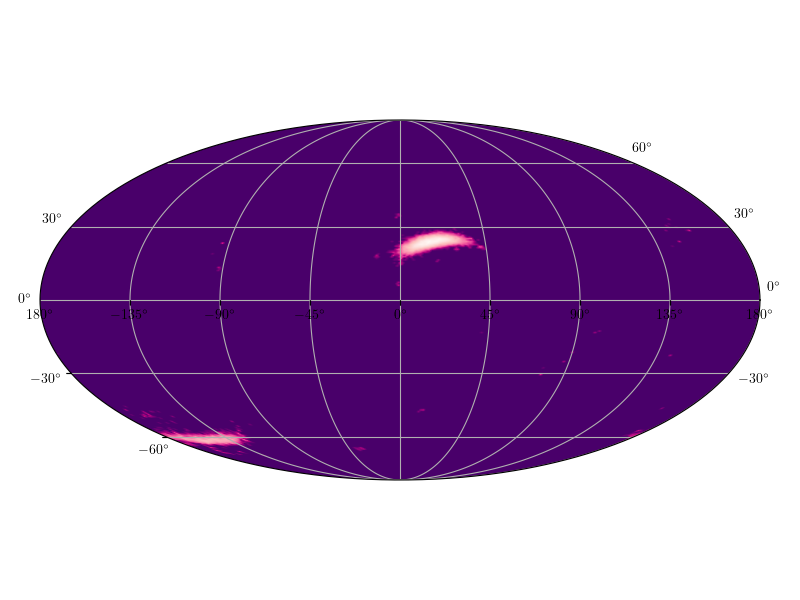}\hfill
    \includegraphics[width=0.31\textwidth]{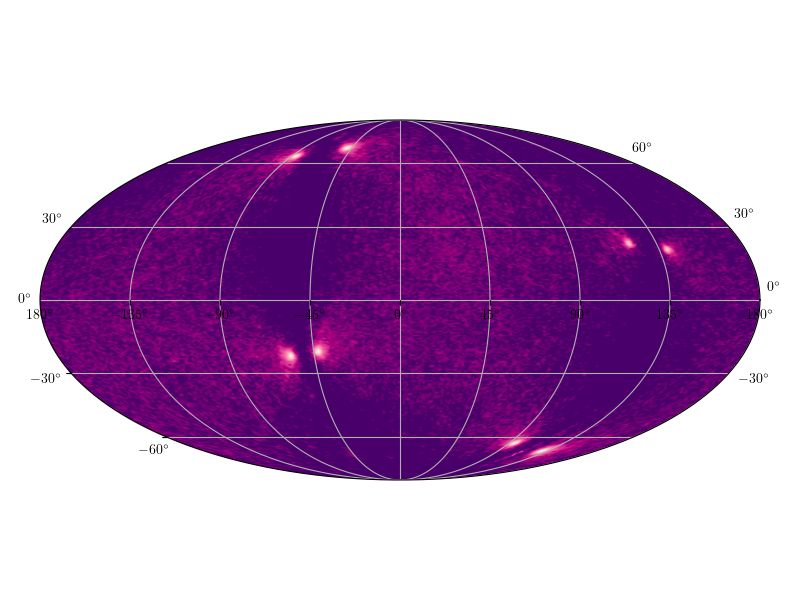}\hfill
    \includegraphics[width=0.31\textwidth]{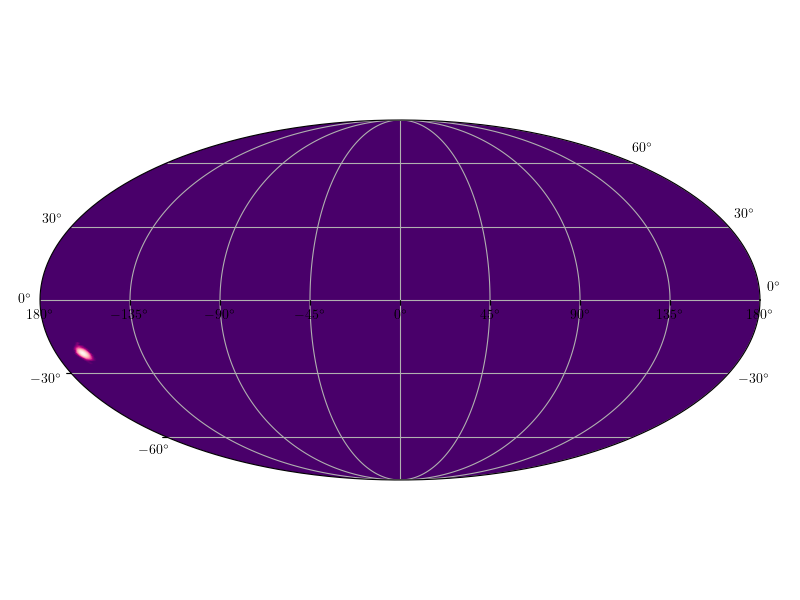}\hfill
    \vspace{-3em}
    \includegraphics[width=0.31\textwidth]{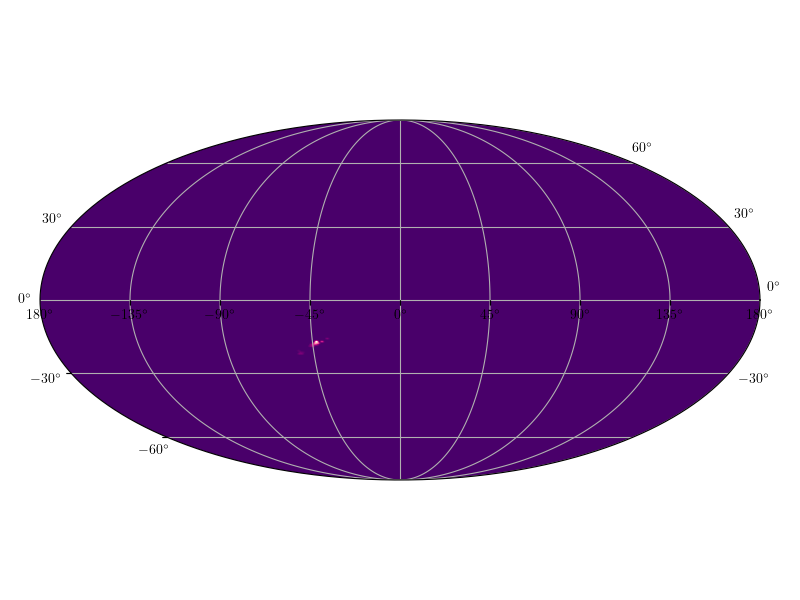}\hfill
    \includegraphics[width=0.31\textwidth]{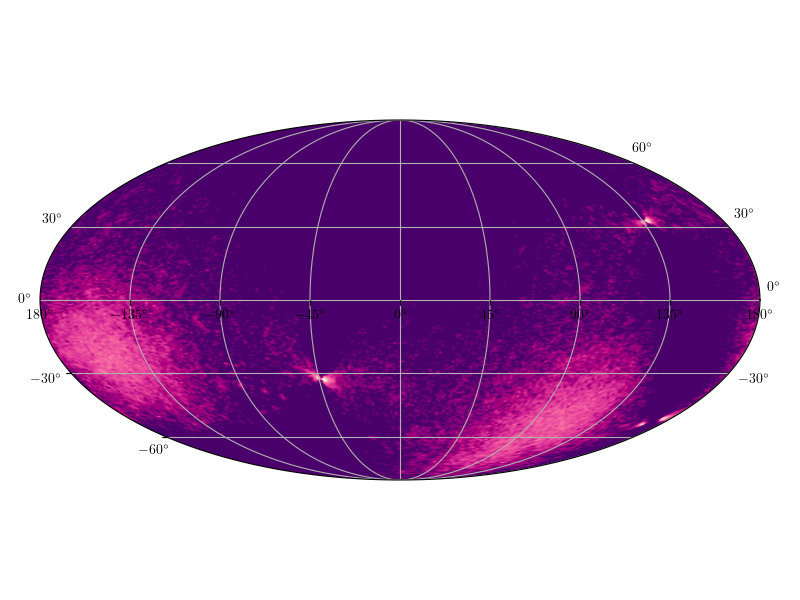}\hfill
    \includegraphics[width=0.31\textwidth]{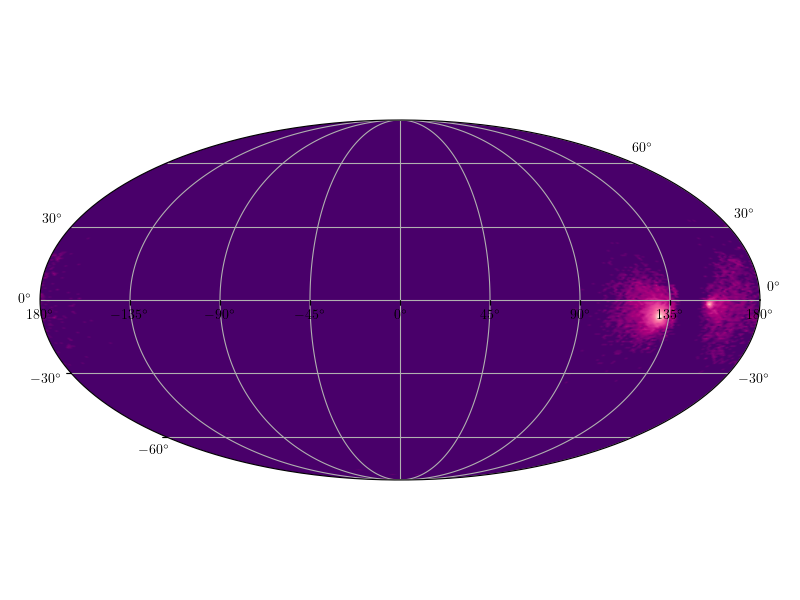}\hfill
    \vspace{-3em}
    \includegraphics[width=0.31\textwidth]{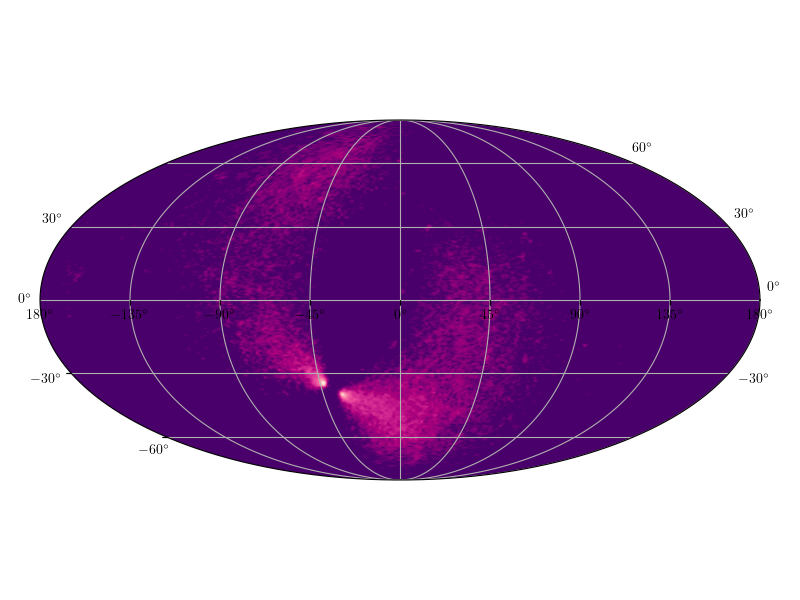}\hfill
    \includegraphics[width=0.31\textwidth]{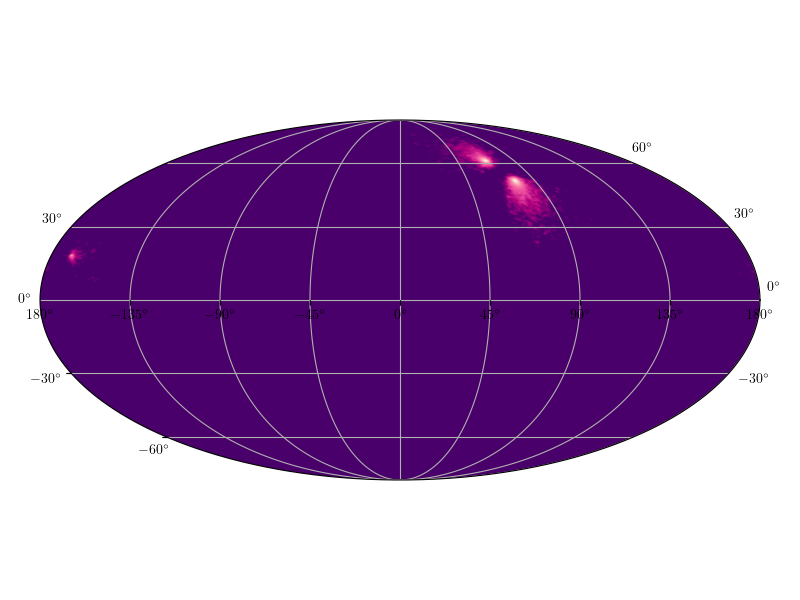}\hfill
    \includegraphics[width=0.31\textwidth]{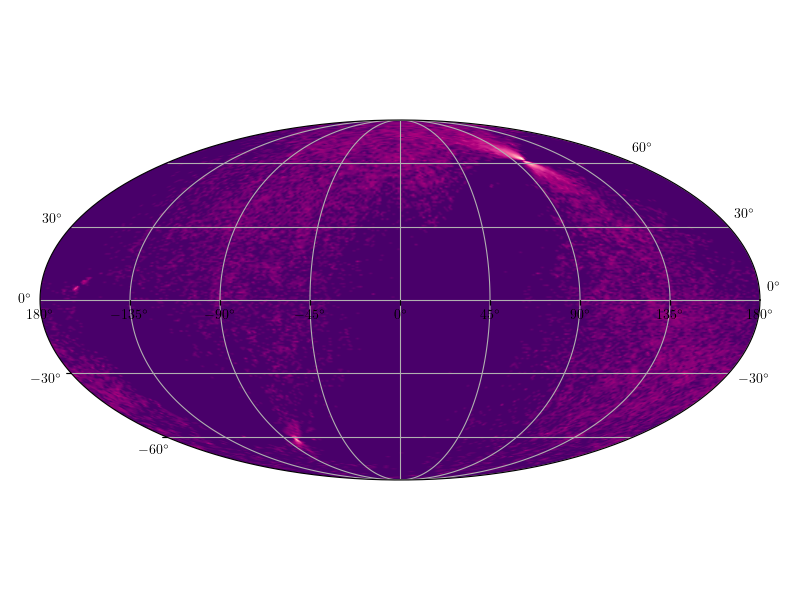}\hfill
    \vspace{-3em}
    \includegraphics[width=0.31\textwidth]{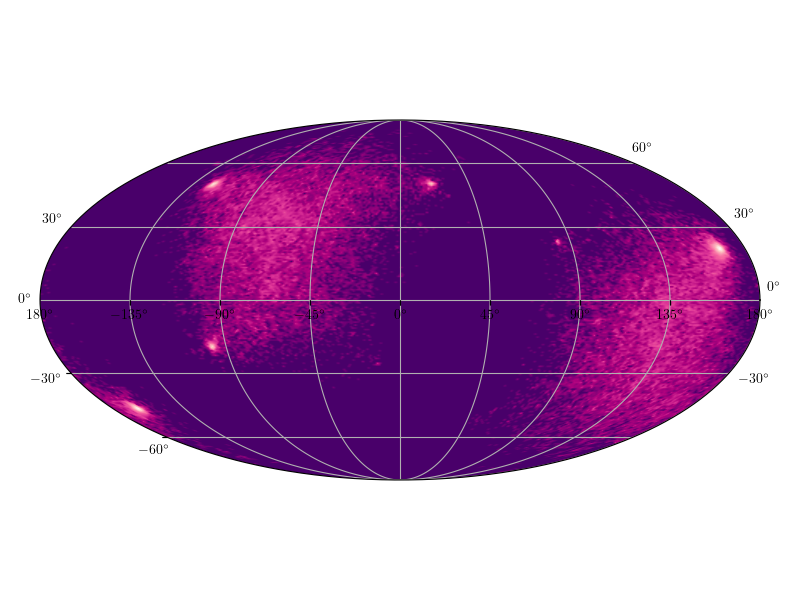}\hfill
    \includegraphics[width=0.31\textwidth]{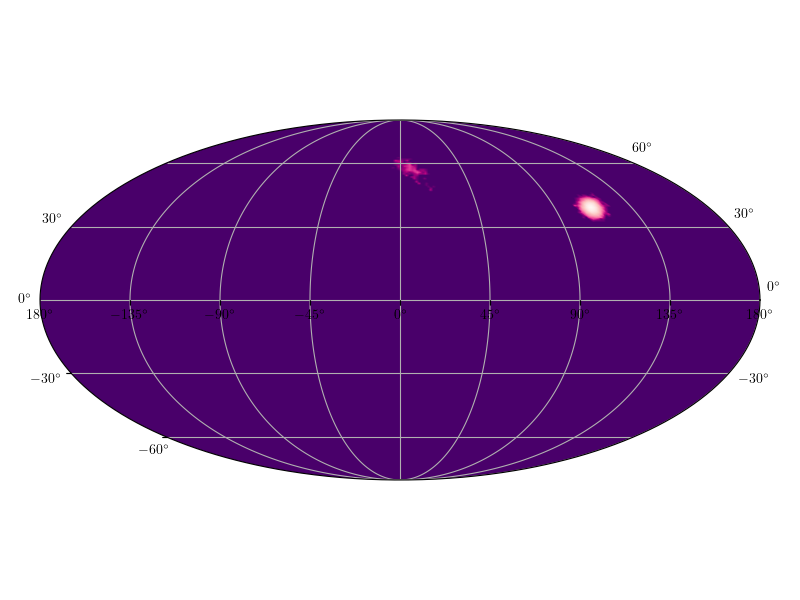}\hfill 
    \includegraphics[width=0.31\textwidth]{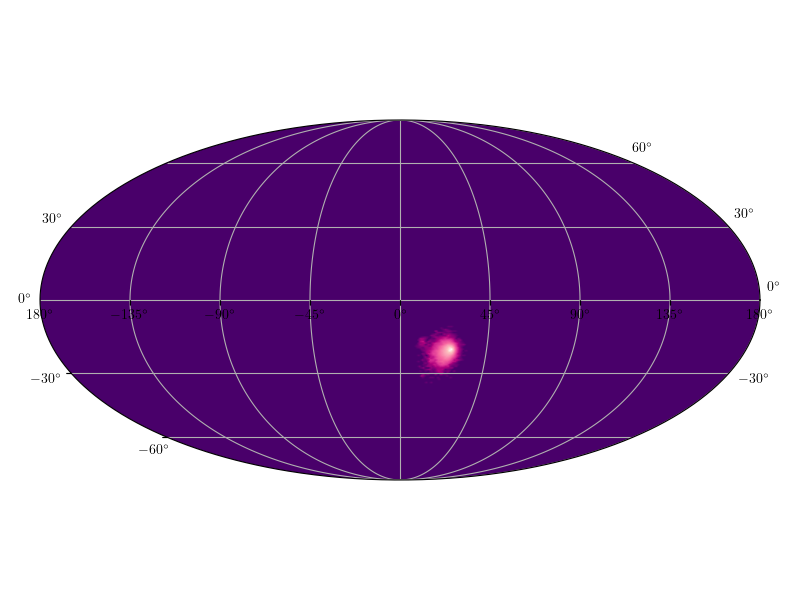}\hfill
    \vspace{-3em}
    \includegraphics[width=0.31\textwidth]{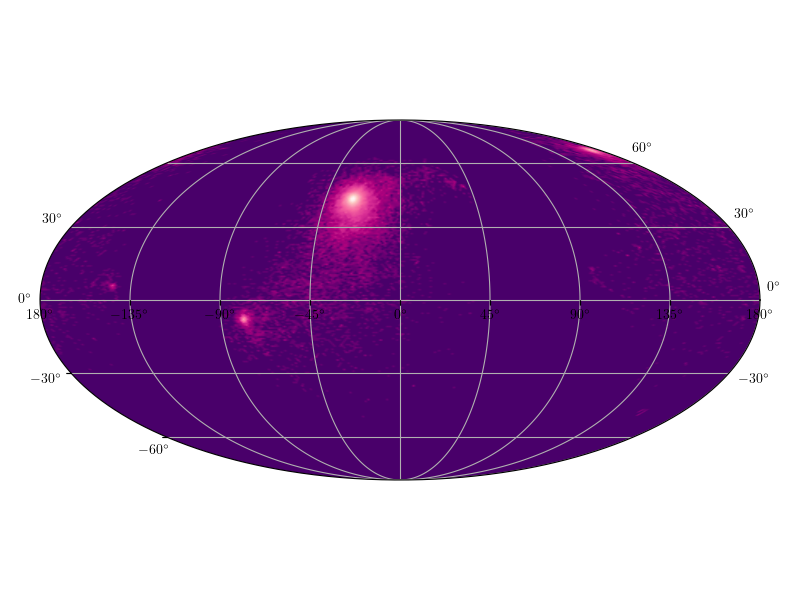}\hfill
    \includegraphics[width=0.31\textwidth]{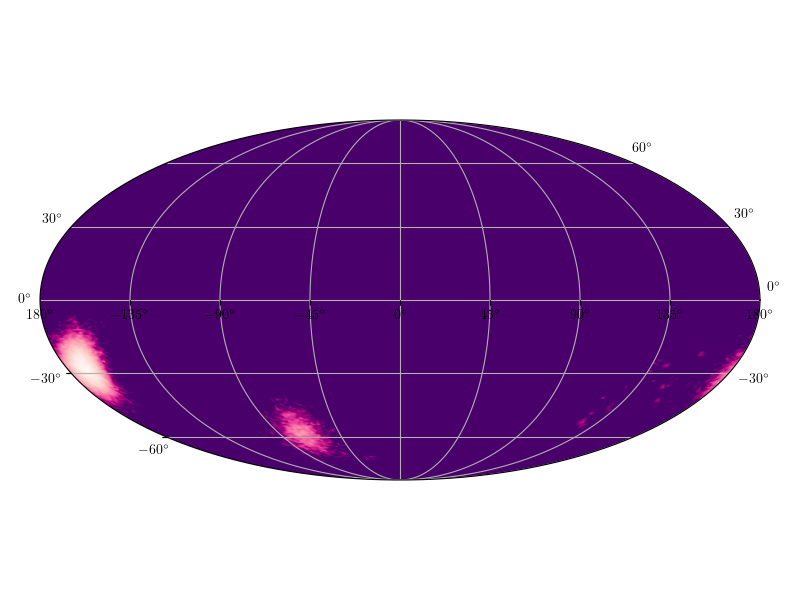}\hfill
    \includegraphics[width=0.31\textwidth]{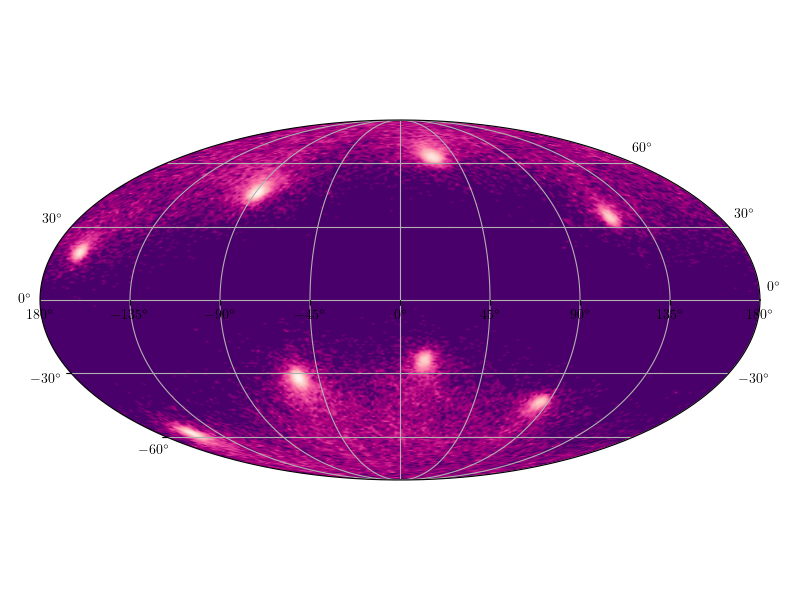}\hfill
    
    \caption{\small Marginalized distributions of MBHB sky location in ecliptic coordinates. The sky maps are ordered by merger time from top left to bottom right. The different morphologies of the sky maps are due to the different maximum frequency and duration of the signals, with lower mass binaries reaching high frequency and spending more time in band leading to precise sky localization. Degeneracies in the sky location improve when higher harmonics are included in the waveforms.}
    \label{fig:mbh_sky_mollweide}
\end{figure*}

For the final demonstration of the MBHB catalog, fig.~\ref{fig:mbh_sky_mollweide} displays the sky map for each MBHB source in the catalog, in order of merger time from top left to bottom right, after the full 12 month analysis.
The variety found in the MBHB sky maps is the result of the relative importance for each event of the three different ``channels'' of localization information for these signals.  
The first channel is through the GW phase which is frequency-modulated by the orbital motion of the spacecraft.  
The second channel of localization information comes from the relative arrival time of the GW signal's wave-fronts at the different spacecraft. 
The third, and least informative, is from the non-uniform detector response over the sky which is encoded in the relative amplitudes of the GW signal in the different TDI channels.
Sky maps that contain many modes of high probability are typically from higher mass binaries that are shorter lived in the LISA data, missing out on the Doppler modulations induced by the orbital motion of the detector and not reaching high enough frequency to differentiate the signal arrival times at the different spacecraft.
Lower mass MBHB signals get the best of both worlds, as they are in the measurement band for long enough to have clearly detectable frequency modulation and they reach short enough wavelengths to benefit from the time of arrival measurements. 
Note that even for the high mass and short-lived binaries additional information from spin precession and, more importantly, higher harmonics of the waveform will help break degeneracies~\cite{Marsat:2020rtl}. 
See Ref.~\cite{PhysRevD.103.083011} for a more comprehensive discussion and demonstration of MBHB parameter estimation with LISA. 

\section{Discussion and Future Work}\label{sec:discussion}
The global analysis demonstrated here is one important step towards a fully functioning pipeline ready for LISA observational data but there is still a long way to go. 
Obviously missing are the other anticipated source types, though the \TheGlobalFit\ architecture is designed to seamlessly accommodate additional modules. 
Extending to other source types is generally expected to be a small perturbation relative to the overall scale of the analysis which is set by the UCBs.

Another obvious direction of development is to reduce the overall computational cost of the analysis. 
The current version of \TheGlobalFit\ used $\mathcal{O}(10^3)$ CPUs for $\mathcal{O}(5)$ days to process the 12 month data, and those processing times will increase roughly linearly as the observation time grows. 
The current algorithm will become uncomfortably expensive for multi-year data sets.
Reducing the computational cost is of crucial importance for further development because the main source of stress on the analysis methods provided by LISA data is the scale.
Optimal development of the global analysis requires frequent processing of full-scale data sets.
The two prongs for reducing the computational cost of the analysis are by lowering the cost of each likelihood evaluation with accelerated compuatational techniques (e.g. Ref~\cite{PhysRevD.102.023033}), and by reducing the total number of likelihood evaluations by developing a more efficient sampling algorithm through further development of data- and domain-driven proposal distributions.

Beyond implementation improvements, there are a number of model assumptions, reflected in the simplicity of the likelihood function, that will need to be relaxed to properly handle observational data. 
Generally speaking, it is the assumption of stationary noise that is most problematic, though there are different sources of non-stationarity that each deserve their own strategy.
As mentioned earlier, there will be periodic (due to the galactic foreground) and secular (due to the instrument) changes in the instrument noise level which introduce non-zero off-diagonal elements of the frequency-domain noise covariance matrix.  
We will mitigate these effects by moving away from conducting the analysis in the Fourier domain, in favor of a discrete wavelet basis, which still yields a diagonal noise covariance matrix if the stationary timescales are longer than the duration of the wavelet basis functions.  
Descriptions of waveform and noise models in the wavelet domain are found in Refs~\cite{PhysRevD.102.124038,Digman:2022igm,Digman_2022} and are scheduled to be integrated into \TheGlobalFit.
In the wavelet domain the noise model will be a function of both time and frequency to track the slow drift in the instrument and foreground noise levels.

The time-frequency approach renders the heterodyning currently used for both the UCB and MBHB likelihood calculations redundant. 
Wavelet decompositions incorporate a natural compression of GW signals since the likelihood only integral only changes along the signal track $f(t)$, which has length ${\sim}\sqrt{N}$ for a data set with $N$ data points~\cite{PhysRevD.102.124038,Digman:2022igm}. 
Wavelet domain likelihoods are typically faster than their heterodyned analogs without requiring a reference waveform or any pre-computation step.
Faster likelihood functions allow for more rapid convergence and better mixing between blocks of the MH sampler for the same computational cost.
The wavelet domain is also better suited to handling data gaps than frequency domain analyses.
In the wavelet domain the basis functions are finite duration with built-in window functions that naturally suppress spectral leakage caused by gaps in the data.
Fourier methods require additional data conditioning to deal with gaps, either through windowing or data augmentation~\cite{PhysRevD.100.022003}.

Another modeling limitation of the current analysis is that it treats the unresolved galactic signals as {\em noise}, when in reality they are better described as a cyclostationary stochastic {\em signal}. Going forward we will explore using a physically parameterized instrument noise model~\cite{Adams:2013qma}, while treating the unresolved galactic binaries as a separate, time-varying, stochastic signal~\cite{Adams:2013qma,Digman_2022}. Such a treatment requires that we include the (approximately, at low frequency) noise-only TDI ``T'' channel, and not just the A and E channels as is done in the current version. 
Modeling of the galactic confusion will be further improved by coupling the resolved UCB population in the global fit to a physical model of the foreground via parameterized priors for the spatial distribution of binaries~\cite{PhysRevD.86.124032} and the overall number density of sources as a function of frequency.

A source of non-stationary noise not well suited for modeling with the power spectral density (or time-frequency equivalent) are short duration noise transients or ``glitches.''
The path to incorporating a glitch model (and, by corollary, a model for generic GW transients) into \TheGlobalFit\ has already been paved by similar work done for LIGO-Virgo data~\cite{PhysRevD.103.044006} and theoretical demonstrations using simulated LISA data~\cite{PhysRevD.99.024019}.
There is already LDC data that contain a simulated glitch population informed by the LISA \emph{Pathfinder} observations~\cite{PhysRevLett.120.061101} and the incorporation of a transient noise module into the \TheGlobalFit\ framework is a near-term priority.

One final currently planned development direction for \TheGlobalFit\ is in the instrument model itself, enabling the global analysis to start with lower level data products than the TDI channels currently used.
A data-driven approach to perform self-calibration coupled with the global fit naturally propagates uncertainties at each stage of the signal processing chain to the astrophysical inferences made with the GW signals. 
Such capabilities may prove to be important for science investigations where control of systematic errors are vital, the most obvious example of which would be testing the nature of gravity with high S/N MBHBs or EMRIs.
Already-demonstrated examples of self-calibration methods for LISA include employing UCBs as phase standards~\cite{PhysRevD.98.043008} and using the phasemeter data to infer the light travel time between spacecraft for cancellation of laser frequency noise and construction of the TDI interferometer combinations~\cite{PhysRevD.71.041101, PhysRevD.104.084037}.
Another possible capability to explore is the removal of noise in the inter- and intra-spacecraft interferometer measurements caused by angular jitter of the test masses masquerading as distance fluctuations--the so-called ``tilt to length coupling'' inherent in the LISA measurement~\cite{Trobs_2018}.
An instrument module in the \TheGlobalFit\ is valuable for quantitatively understanding and, if necessary, mitigating the affect of calibration uncertainties an astrophysical inferences.

As is clear from the long list of future work, the \TheGlobalFit\ architecture described in this paper does not represent a finished design but instead is the scaffolding upon which further development will be built.
Nevertheless, our demonstrated results are an important way-point on the path towards a fully functional pipeline ready for LISA observational data.

\section{Acknowledgements}

\emph{Software:} Results presented here used v2.0 of \hyperlink{github.com/tlittenberg/ldasoft}{\tt ldasoft}, a public C library which includes the noise, UCB, VGB, and global fit samplers. The MBHB sampler is managed independently at \hyperlink{github.com/eXtremeGravityInstitute/LISA-Massive-Black-Hole}{\tt LISA-Massive-Black-Hole}. Postprocessing and visualization tools for the source catalogs are available the python package \hyperlink{github.com/tlittenberg/lisacattools}{\tt lisacattools} which in turn depends on {\tt numpy}~\cite{harris2020array}, {\tt pandas}~\cite{reback2020pandas,mckinney-proc-scipy-2010}, {\tt matplotlib}~\cite{Hunter:2007}, {\tt astropy}~\cite{2022ApJ...935..167A}, {\tt seaborn}~\cite{Waskom2021}, and {\tt ChainConsumer}~\cite{Hinton2016}.

The authors thank K. Gresbach for multithreading the \gbmcmc\ pipeline; J. Baker, K. Lackeos, T. Robson, J. Slutsky, and J.I. Thorpe for their useful discussions and suggestions during the development of the global fit pipeline and the assessment of the results; J.C. Malapert and J.I. Thorpe for their role as co-developers of {\tt lisacattools}; J.H. Thompson for indispensable help deploying the pipeline on the AWS resources; and the LISA Data Challenge group for providing and supporting the simulated data. T.B. Littenberg is supported by the NASA LISA Study Office. N.J. Cornish appreciates the support of the NASA LISA Preparatory Science Grant 80NSSC19K0320.

\bibliography{references}

\end{document}